\documentclass[journal]{IEEEtran}
\usepackage[utf8]{inputenc}
\usepackage[english]{babel}
\usepackage{multirow}
\usepackage{graphicx}
\graphicspath{ {./images/} }
\usepackage{booktabs}
\usepackage{graphicx}
\usepackage{tabularx}
\usepackage{etoolbox}
\usepackage{nomencl}
\usepackage{breqn}
\usepackage[linesnumbered,ruled,vlined]{algorithm2e}
\usepackage{tikz}
\usepackage{array}
\makenomenclature
\usepackage{amsfonts}
\usepackage{adjustbox}
\usepackage{breqn}
\usepackage{color}
\usepackage{xcolor}
\usepackage{makecell}
\usepackage{amsmath}
\usepackage{nicematrix}
\usepackage{cite}
\usepackage{textcomp}
\usepackage{stfloats}
\usepackage{url}
\usepackage{verbatim}
\usepackage{algorithmicx}
\usepackage{cite}
\usepackage{relsize}
\usepackage[caption=false,font=normalsize,labelfont=sf,textfont=sf]{subfig}
\SetKwInput{KwInput}{Input}                
\SetKwInput{KwOutput}{Output}   
\usepackage{hyphenat}
\usepackage[hidelinks]{hyperref}
\hypersetup{
    colorlinks=false,
    linkcolor=black,
    filecolor=magenta,      
    urlcolor=cyan,
    pdftitle={Overleaf Example},
    pdfpagemode=FullScreen,
    }
\usepackage{comment}
\usepackage{mathrsfs}
\usepackage{booktabs}
\usepackage{svg}
\usepackage{setspace}
\usepackage{enumitem}
\usepackage{dsfont}
\usepackage{pict2e}

\allowdisplaybreaks
\newif\ifarxiv
\arxivtrue

\vspace{-3em}

\title{Co-Optimization of Damage Assessment and Restoration: A Resilience-Driven Dynamic Crew Allocation for Power Distribution Systems}

\author{Ali Jalilian, Babak Taheri,
and Daniel K. Molzahn
        \thanks{A. Jalilian is with the Electrical Engineering Department, Sharif University of Technology, Tehran, Iran. Email: {\tt\small{alijalilian918@yahoo.com}}.

       B. Taheri and D.K. Molzahn are with the School of Electrical and Computer Engineering,
        Georgia Institute of Technology, Atlanta, GA 30332, USA.
        Emails: {\tt\small{\{taheri, molzahn\}@gatech.edu}}. This research was supported by NSF award \#2145564.
        }
        \vspace{-1em}

}

\begin{document}
\maketitle
\begin{abstract}
This study introduces a mixed-integer linear programming (MILP) model, effectively co-optimizing patrolling, damage assessment, fault isolation, repair, and load re-energization processes. The model is designed to solve a vital operational conundrum: deciding between further network exploration to obtain more comprehensive data or addressing the repair of already identified faults. As information on the fault location and repair timelines becomes available, the model allows for dynamic adaptation of crew dispatch decisions. In addition, this study proposes a conservative power flow constraint set that considers two network loading scenarios within the final network configuration. This approach results in the determination of an upper and a lower bound for node voltage levels and an upper bound for power line flows. To underscore the practicality and scalability of the proposed model, we have demonstrated its application using \texttt{IEEE 123-node} and \texttt{8500-node} test systems, where it delivered promising results.
\end{abstract}



\section*{Nomenclature}

\begin{spacing}{0.5}
\textbf{Sets and Indexes:}
\begin{align*}
\mathcal{B}, b & \quad \text{Set and index of buses} \\
\mathcal{L}, \ell & \quad \text{Set and index of sections (lines)} \\
\mathcal{Z}, z & \quad \text{Set and index of electrical zones} \\
\mathcal{Q}, q & \quad \text{Set and index of unpatrolled zones} \\
\mathcal{R}, r & \quad \text{Set and index of RCSs} \\
\mathcal{M}, m & \quad \text{Set and index for manual switches (MS)} \\
\mathcal{F}, f & \quad \text{Set and index of faults} \\
\mathcal{C}, c & \quad \text{Set and index of available crews} \\
\mathcal{E}, e & \quad \text{Set and index of equipment in patrol zones} \\
\mathcal{P}, p & \quad \text{Set and index of all locations in crew routing} \\
\mathcal{T}, t & \quad \text{Set and index of time steps} \\    
\end{align*}

\vspace{-0.75em}
\textbf{Subsets:}
\begin{align*}
\mathcal{M}\backslash\mathcal{R}_{z, z^{\prime}} & \quad \text{Set of MSs $\backslash$ RCSs connecting $z$ and $z^{\prime}$} \\
\mathcal{F}_{z} \backslash \mathcal{B}_{z} & \quad \text{Set of faults $\backslash$ buses in $z$} \\
\mathcal{P}_{\mathcal{C}} & \quad \text{Set of crews' initial locations} \\
\mathcal{P}_{\mathcal{F}} \backslash \mathcal{P}_{\mathcal{M}} & \quad \text{Set of faults $\backslash$ MSs' locations} \\
\mathcal{P}_{\mathcal{M}^{\prime}} & \quad \text{Duplicate set of MSs' location for 2nd switching} \\
\mathcal{F}_{\mathcal{Q}} & \quad \text{Set of hypothetical faults in unpatrolled zones} \\
\end{align*}

\vspace{-0.75em}
\textbf{Parameters:}
\begin{align*}
T^{\text {repair }} & \quad \text {Required repair time for faults} \\
T^{\text {patrol }} & \quad \text {Estimated patrol time of patrol zones} \\
\rho_{e} & \quad \text {Failure probability of equipment} \\
C_{z}^{\text {out }} & \quad \text {Cost coefficient commensurate to ENS} \\
C^{\text {tra }} & \quad \text {Cost coefficient commensurate to crews' travels} \\
\Delta_{p, p^{\prime}} & \quad \text {Travel time between two points for crews} \\
BT & \quad \text {A large out-of-scope amount of time} \\
P_{z} & \quad \text {Zonal power consumption} \\
M & \quad \text {Big-enough constant positive value} \\
\alpha_{b}^{\text {sub }} & \quad \text {Binary value showing if a bus is a substation} \\
\beta_{m}^{\text {MSI }} & \quad \text {Binary value showing if an MS is initially closed} \\
D_{b} & \quad \text {Active and reactive demand} \\
\end{align*}

\vspace{-0.75em}
\textbf{Binary Variables:}
\begin{align*}
\beta_{p, p^{\prime}} & \quad =1 \text{ if a path from $p$ to $p^{\prime}$ is traversed by a crew} \\
\beta_{p}^{V} & \quad \text{Indicates if a crew visits $p$} \\
\beta_{m}^{MSP} & \quad \text{Indicates if a crew opens an MS during a patrol} \\
\beta_{z^{\prime}, z}^{zz} & \quad \text{Indicates if $z$ is energized by $z^{\prime}$} \\
\beta_{m}^{MSF} & \quad \text{Indicates if an MS is finally closed} \\
\beta_{r}^{RCS} & \quad \text{Indicates if an RCS is finally closed} \\
\beta_{\ell}^{\text {line }} & \quad \text{Indicates if a line is finally connected} \\
\alpha_{z}^{\text {root }} & \quad \text{Indicates if there is a substation or a master DG} \\
\alpha_{b}^{DG} & \quad \text{Indicates if there is a master DG in $b$} \\
\beta_{z, t}^{zt} & \quad \text{Indicates if a zone is energized in a time step} \\
\zeta_{z, z^{\prime}} & \quad \text{Indicates if zone $z$ is energized earlier than $z^{\prime}$} \\
\end{align*}

\vspace{-0.75em}
\textbf{Continuous Variables:}
\begin{align*}
T_{z}^{\text {out }} & \quad \text {Outage time} \\
\tau_{p}^{c} & \quad \text {Finish time of an action in } p \text { by crews} \\
T_{p}^{\text {op }} & \quad \text {Operation time for a remedial action in } p \\
U_{b} & \quad \text {Voltage magnitude of buses} \\
\varphi_{\ell} \backslash G_{b} & \quad \text {Active and reactive line flow $\backslash$ power generation}\\
\end{align*}

\end{spacing}


\vspace{-1em}
\section{Introduction}\label{introduction}

\IEEEPARstart{C}{ritical} infrastructures (CIs), such as electricity, are integral to the functioning of societies. These backbones of economy, security, and health are increasingly susceptible to high-impact, low-probability (HILP) events, including natural disasters and adverse weather conditions \cite{Panteli2015, DOH}. A disruption in these infrastructures, especially in power distribution systems, not only affects other essential CIs, like transportation, communication, and water supply, but also has considerable societal consequences.
With climate change intensifying the frequency and severity of such extreme events, the resilience of power systems, i.e., their ability to prepare for, withstand, and recover swiftly from disruptive events, is gaining increased attention. Traditional power systems designed to endure low-impact high-probability (LIHP) events are being challenged to evolve and handle these significant HILP incidents. The need for resilience is particularly critical at the distribution level, where $80$ -- $90\%$ of power outages occur \cite{Billinton}, thus justifying the recent surge in related research.

This paper addresses this critical issue, focusing on strategies to expedite power restoration following disruptions at the distribution level. It offers a comprehensive model that takes into account fault isolation, damage assessment, network reconfiguration, and microgrid formation. Our model aims to bridge gaps in existing literature, particularly in dealing with these complex, interrelated processes.
Therefore, our literature review touches upon five pivotal facets in the realm of power system restoration: micro-grid formation, network reconfiguration, fault isolation, damage assessment, and addressing technical constraints.

\textit{\textbf{Microgrid Formation}}: As access to the upstream network is often impaired during fault conditions, deploying a multitude of distributed energy sources at the distribution network level in a microgrid can improve resilience. Studies \cite{li2021, sun2022} have emphasized the importance of such resources in the form of distributed generators (DGs) or mobile energy units \cite{taheri2023improving}. While a substantial amount of research has focused on the energy sufficiency, economic viability, and technical limitations of microgrids, others have shed light on microgrid formation through network reconfiguration tactics \cite{cai2021, macedo2021}.

\textit{\textbf{Network Reconfiguration}}: A multi-stage load restoration process inherently calls for iterative network reconfigurations, utilizing sectionalizing switches at each stage. These switches could be remote-controlled or manual. The act of manual switching necessitates field crew presence, which could extend the switching time due to variables such as geographical attributes, traffic conditions, and crew availability. Various studies have dissected the implications of the remote-controlled switches' (RCS) switching actions in distribution networks \cite{ye2020, jiang2022}. Manual switches (MSs), i.e. manual sectionalizers, cut-out fuses, or even circuit breakers without remote control capability, also provide pragmatic and efficient load restoration capabilities. Also, due to the possibility of damage to the cyber network, especially in the event of severe fault conditions \cite{tian2022}, remotely unreachable RCSs could still be engaged manually to help achieve a faster restoration. However, few references have incorporated the optimal performance of MSs in their proposed restoration processes. In \cite{chen2018, zhang2020, liu2022}, operation crews for closing MSs were considered. These papers assume that all of the MSs have been opened in the fault isolation phase. This assumption overlooks the importance of optimal fault isolation.

\textit{\textbf{Fault Isolation}}: Establishing optimal primary fault isolation paves the way for accelerated load pick-up during the restoration process. However, this crucial step has been overlooked in several studies \cite{li2021, sun2022, taheri2023improving, cai2021, macedo2021, ye2020, jiang2022, tian2022, chen2018, zhang2020, liu2022}. On the other hand, other strategies, such as the minimum-area fault isolation approach proposed in \cite{yang2022}, target the isolation of faults through the strategic opening of the nearest switches.  A similar approach in \cite{li2021J} performs fault isolation by disconnecting the predefined set of upstream and downstream sectionalizers. The fault isolation scheme in \cite{arif2019} enforces a zero voltage for terminal buses of a faulty line. Then optimal RCS switching ceases zero-voltage propagation. Reference~\cite{bian2022} considers fault isolation optimization by determining MSs' optimal open/close operation. Therefore, two-time manual switching is incorporated into their crew dispatch.

\textit{\textbf{Damage Assessment}}: A recurrent assumption in outage management research stipulates the known parameters of damage locations and repair times \cite{van2015transmission, rhodes2021powermodelsrestoration, vanhentenryck2011}. However, such an assumption has been challenged in some references \cite{arif2019}. For instance, \cite{sun2022} proposed a dynamic crew grouping dispatch algorithm to overcome the unpredictability of repair workloads while still assuming awareness of the faults' locations from an already finished assessment phase. In the recommended coordinated damage assessment with service restoration scheme in \cite{bian2022}, the repair times and locations of faults are dynamically prepared by assessors and fed to the restoration model. However, the assessment process itself is not optimized. In \cite{liu2021}, fault location, fault isolation, and service restoration for healthy parts of the network are coordinated. However, the proposed method does not involve infrastructure repair. \textcolor{black}{Damage assessment provides two important parameters: fault locations and fault repair times. The unavailability of such information imposes challenges when optimizing the fault management process since the data collection process is time consuming and can interfere with other field corrective actions (switching and repair). Therefore, an integrated framework incorporating various tasks of fault location and damage assessment via feeder patrolling, fault repair, reconfiguration, and load recovery is of paramount importance.}

\textit{\textbf{Technical Constraints}}: 
{Some studies have employed fixed-time steps with a single set of power flow equations for each step \cite{yang2022, arif2019}. However, this methodology can compromise the solution's optimality and escalate computational complexity \cite{chen2018}. Interestingly, \cite{chen2018a} posits that checking power flow solely in a network's final configuration might be sufficient to ensure the safe operation of preceding configurations. This claim, however, comes with caveats. For instance, even though distributed generators (DGs) are factored into the model, zones with DGs must not connect either to one another or to a substation. To avoid explicit power flow models at every juncture, the authors in \cite{chen2018a} introduce specific checkpoints. These checkpoints accommodate configurations that feature multiple connected zones with DGs or a substation. Notably, the count of these checkpoints correlates with network elements, such as DGs or capacitors, that can potentially raise voltage levels.}


Despite the fact that numerous research efforts have tackled different aspects of power service restoration, none has introduced an efficient model that optimizes the process holistically. Given the interdependency of different restoration stages, strategies, and limitations, there is an urgent need for a comprehensive model that is both inclusive and computationally efficient enough for real-world networks. We propose a model that addresses these shortcomings, integrating different stages into a unified restoration package using a state-of-the-art approach. The main contributions of our model include:

\begin{itemize}
\item
Streamlining the entire restoration process, our model integrates everything from damage assessment and fault isolation to repair and network re-energization, thereby avoiding decision-making conflicts in resource allocation for different outage management tasks. \textcolor{black}{We emphasize that, to the best of our knowledge, none of the existing literature holistically considers these aspects when optimizing restoration processes.}
\item
Uncovering non-anticipative information about fault locations and repair times, our model incorporates an ongoing damage assessment. A dynamic proactive-responsive re-optimization framework is deployed for this process.
\item
Improving the description of manual switching throughout the entire process, our methodology takes into account both during-patrol and after-patrol open/close actions.
\item
Incorporating two network loading conditions into conventional power flow constraints at the network's final configuration, our model establishes boundaries for nodes' voltage levels and limits for line power flows. This results in safe operations across all stages, and importantly, the proposed constraints do not necessitate the segregation of zones with power sources, such as substations or DGs.
\item
Demonstrating the effectiveness and scalability of our proposed algorithm through numerical experiments, we present results from medium- and large-scale test cases.
\end{itemize}

This paper is structured as follows: Section~\ref{sec:proposed-methodology} explains our proposed methodology. Section~\ref{sec: Numerical Results} shows our numerical results. Section~\ref{sec: Conclusion} offers conclusions and future directions.

\section{Proposed Methodology}\label{sec:proposed-methodology}
Our methodology devises an intelligent decision-making framework tailored for the complex process of restoring a distribution network after severe weather-induced equipment failures. By balancing system repair tasks, switching operations, and damage assessments, this methodology navigates the challenges efficiently.

\vspace{-1.5em}
\subsection{Decision Framework}\label{decision-framework}
\subsubsection{Event Description and Network Blackout}
Severe weather is notorious for instigating a chain of equipment failures within distribution networks. Protective devices, sensing these faults, trigger automatic shutdown protocols in the preliminary phase of such an event. The situation often exacerbates as the event unfolds, causing more damage, and inducing more faults. \textcolor{black}{In this condition, multiple equipment failures, communication outages, or even the power outage itself may limit situational awareness about the network.} Field crews can only be deployed once safe operational conditions are restored. As a result, during this stage, comprehensive information regarding the damage—such as the number and location of faults, extent of the damage, and anticipated repair duration—is typically scarce.

\subsubsection{Damage Assessment and Patrol Tasks}
In light of the transportation network's characteristics, the distribution feeder is divided into several patrolling areas for damage evaluation and data gathering. Here, we assume that the number and extent of patrol zones are predetermined. Taking into account the event's severity and the equipment's fragility curves, we determine the likelihood of equipment failure \cite{panteli2017}. To each area, we assign a hypothetical fault with a repair time equivalent to the sum of the patrolling duration for that area and the expected repair time. This repair time is deduced from equation \eqref{eq: time}:
\begin{equation}
\label{eq: time}
T_{q}^{\text {repair }} = T_{q}^{\text {patrol }} + \sum_{e \in \mathcal{E}{q}} T_{e}^{\text {repair }} \rho_{e}.
\end{equation}
Here, \eqref{eq: time} computes the repair time $T_{q}^{\text {repair }}$ for the hypothetical fault in patrolling area $q$. This time is the sum of the patrol time $T_{q}^{\text {patrol }}$ and the product of each equipment's repair time $T_{e}^{\text {repair }}$ and failure probability $\rho_{e}$ within the area $\mathcal{E}_{q}$.

\subsubsection{Task Assignment} \label{task assignment}
\begin{figure}
\vspace{-1em}
\centering
\includegraphics[width=2.9in]{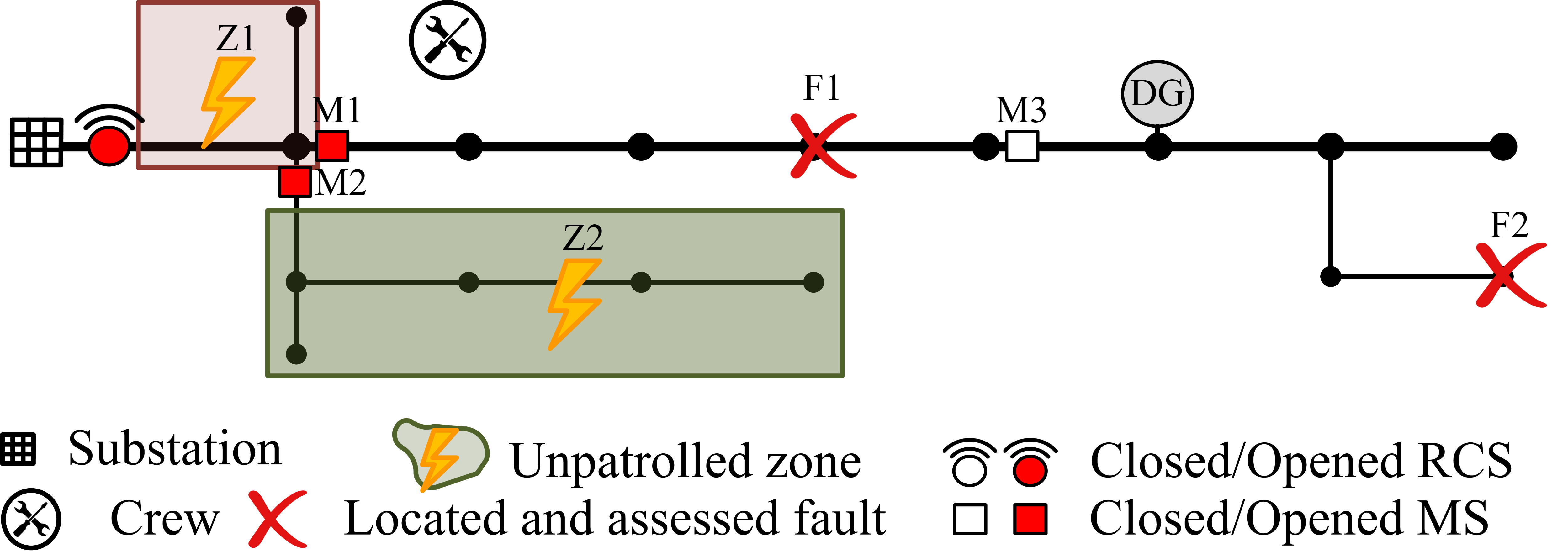}

\caption{Task distribution for repair crews}
\label{fig: repair crews}
\vspace{-1em}
\end{figure}

One of the significant challenges during the restoration process is to determine the optimal allocation of various tasks—such as switching operations and repair of actual and hypothetical (patrolling) faults—to the repair crews. The distribution of tasks among repair crews is depicted in Fig.~\ref{fig: repair crews}.

We consider three types of manual switching actions:
\begin{enumerate}[leftmargin=*]
\item During-patrol MS opening (optimal primary fault isolation).
\item Deploying a crew for the first switching action of an MS (open/close).
\item Deploying a crew for the second switching action of an MS (close).
\end{enumerate}
The first switching type is described within a patrol action, forming a single patrol-and-switch task, while the second and third switching types are single-task duties. Consequently, under our proposed methodology, normally closed MSs can be opened either during patrol or by directly deploying a crew. If an MS is opened, it can be closed via the second switching action. Conversely, normally open switches can only be closed through a first-time direct switching operation. To manage the modeling complexity and computational challenges, in this paper, we do not operate each MS more than two times. This modeling choice prevents reconfiguration of the energized parts of the network in each set of decisions. 

\subsubsection{Chronological Description}

\begin{figure}[h]
\vspace{-1em}
\centering
\includegraphics[width=2.9in]{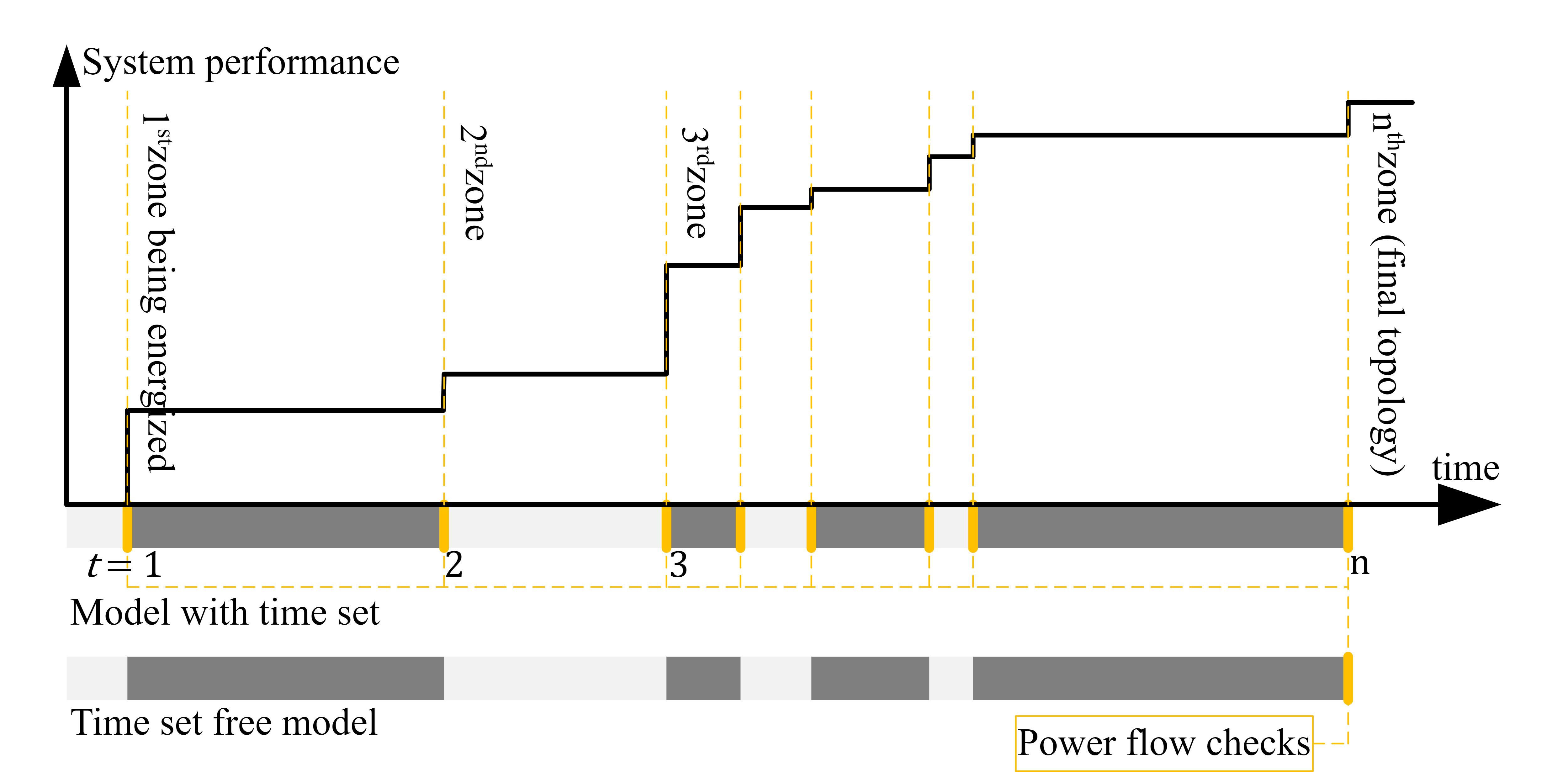}
\caption{Chronological description of the model}
\label{fig: Chronological description of the model}
\vspace{-1.8em}
\end{figure}

{As highlighted in Section~\ref{mathematical-formulation}, we dispatch our crews based on specific routing decisions. Keeping these decisions updated is of utmost importance. To address this, we incorporate proactive re-optimization, scheduled either at set times or regular intervals. Additionally, we use responsive re-optimization, which is initiated after an area has been patrolled or when a new fault comes to light and has been thoroughly evaluated.
For the purposes of our research, we focus on the timings associated with crew actions and the energization of zones. These have been integrated as decision variables within our optimization framework. Consequently, our proposed methodology operates by responding to variable time events. We define a set of events, represented by $ \mathcal{T} $, which captures the order of zone energizations and assigns a unique set of power flow constraints for each instance of $ t \in \mathcal{T} $.
Instead of continually checking, we also consider an alternative approach that verifies power flow only when the network reaches its final configuration, i.e., no further zones are left to be restored. This approach eliminates the need to explicitly model power flow across the expansive set of events, $\mathcal{T}$, thus greatly improving computational efficiency. Fig.~\ref{fig: Chronological description of the model} provides a chronological overview of our model.}

\subsection{Mathematical Formulation}\label{mathematical-formulation}

The primary goal of a restoration plan is to minimize the overall cost incurred from an event. A significant portion of this cost accrues from electric service disruptions. There are also costs associated with the restoration process, such as crew mobilization expenses, which are comparatively minimal but essential to consider to prevent the dispatch of remote crews for certain tasks. The proposed model, grounded in this concept, aims to minimize the total cost:
\begin{equation}
\label{Objective}
\text{Cost} = \sum_{z \in \mathcal{Z}} T_{z}^{\text {out }} P_{z} C_{z}^{\text {out }} + \sum_{p, p^{\prime} \in \mathcal{P}} \beta_{p, p^{\prime}} \Delta_{p, p^{\prime}} C^{\text {tra }},
\end{equation}
where $\mathcal{Z}$ represents the set of all electrical zones, with $z$ as an index. The outage duration is represented by $T_{z}^{\text {out }}$, $P_{z}$ is the power consumption, and $C_{z}^{\text {out }}$ is a cost coefficient corresponding to the energy not supplied. The first term represents the customers' damage costs, which is a function of these variables. 
In the second term, $\mathcal{P}$ denotes the set of all locations within the crew routing, with a pair of indexes $(p, p^{\prime})$. The binary variable $\beta_{p, p^{\prime}}$ indicates whether a crew traverses a path from location $p$ to location $p^{\prime}$, $\Delta_{p, p^{\prime}}$ represents the travel time between these locations, and $C^{\text {tra }}$ is a cost coefficient corresponding to the crews' travel. The second term encapsulates the cost associated with crew teams and their vehicles, accounting for the distance covered, the duration of travel, and the related cost coefficient. The summation is performed over all location pairs.

\begin{figure}[h]
\vspace{-1em}
\centering
\includegraphics[width=2.9in]{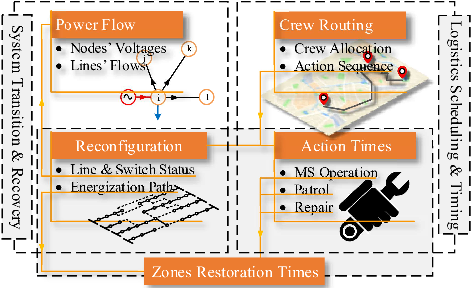}
\caption{High-level description of the constraints}
\label{fig: High level description}
\vspace{-1em}
\end{figure}

The optimization problem we address is bound by multiple technical and operational constraints. Fig.~\ref{fig: High level description} illustrates the primary characteristics of five distinct constraint classes and the interrelationships among them. Notably, action sequences, which are pivotal decision variables in crew routing constraints, have a significant influence over various action timings. This is because an action's completion time is contingent on its placement within a crew’s list of duties. Furthermore, these sequences are crucial for network reconfiguration, as they dictate decisions regarding the switching of MSs.
Each class of constraints will be detailed in the ensuing sections.

\subsubsection{Crew Routing} The process of optimally allocating repair crews for manual switching and fault repairs is a routing problem. As previously discussed, to maintain an accurate description of the restoration process without assuming the MSs are open at the start of the switching process, it is necessary to consider the possibility of two switching operations for each MS. With this in mind, the crew routing constraints are:
\begin{subequations}
\begin{align}
& \sum_{p^{\prime} \in \mathcal{P}} \beta_{p, p^{\prime}} \leq \beta_{p}^{V} ; \quad  \forall p \in \mathcal{P} \label{eq: crew routing:1} \\
& \sum_{p^{\prime} \in \mathcal{P}} \beta_{p^{\prime}, p}=\beta_{p}^{V} ; \quad \forall p \in \mathcal{P} \backslash \mathcal{P}_{\mathcal{C}} \label{eq: crew routing:2} \\
& \beta_{p^{\prime}}^{V} \leq \beta_{m}^{MSP}+\beta_{p}^{V} \leq 1; \nonumber \\  
& \qquad \qquad  \forall m \equiv p \equiv p^{\prime}, m \in \mathcal{M}, p \in \mathcal{P}_{\mathcal{M}}, p^{\prime} \in \mathcal{P}_{\mathcal{M}^{\prime}} \label{eq: crew routing:3} \\
& \beta_{p^{\prime}, p}=0 ; \quad \forall\left(p \in \mathcal{P}_{\mathcal{C}}, p^{\prime} \in \mathcal{P}\right) \text { or } \left(p=p^{\prime} \in \mathcal{P}\right).  \label{eq: crew routing:4}
\end{align}
\label{eq: crew routing}
\end{subequations}
In these equations, $\mathcal{P}_{\mathcal{C}}$ denotes the set of crews' initial locations, and $\mathcal{M}$ represents all MSs. The sets of locations for the first and second switch operations of MSs are given by $\mathcal{P}_{\mathcal{M}}$ and $\mathcal{P}_{\mathcal{M}^{\prime}}$, respectively. \textcolor{black}{The expression $p\equiv {p'}$ signifies that $p$ and $p'$ are pointing to the same location but $p\in\mathcal{P_M}$ and $p'\in\mathcal{P_{M'}}$, i.e., $p$ refers to a manual switching operation occurring for the first time, while $p'$ refers to a manual switching operation of the same switch occurring for the second time.} The binary variable $\beta_{p}^{V}$ indicates whether a crew visits location $p$, and $\beta_{m}^{MSP}$ represents whether a crew operates MS $m$ during a patrol.
Equation \eqref{eq: crew routing:1} states that a crew can only be dispatched from a location if it has been visited. When a remedial action, such as repair or switching, is implemented at a location, \eqref{eq: crew routing:2} ensures a crew is dispatched to that location. It is important to note that the visiting variable $\left(\beta_{p}^{V}\right)$ is set to $1$ for the crews' initial locations and fault locations since the restoration horizon includes repairing all faults and energizing all loads. Furthermore, \eqref{eq: crew routing:3} allows a second switching of an MS if a crew has been directly dispatched to an MS for the first switching or the MS was opened during a patrol operation. Finally, \eqref{eq: crew routing:4} restricts the route selection variable \( \beta_{p^{\prime}, p} \). Specifically, this variable is set to zero when: (i) the route destination is the crews' initial locations\textcolor{black}{, given that all manual operation locations are defined outside $\mathcal{P}_{\mathcal{C}}$}; (ii) the starting and destination locations are identical. \textcolor{black}{It is important to note that the last condition does not necessarily imply that a crew must move after every re-optimization. Consider a scenario where a crew is currently engaged in a task, such as repairing a fault, when a decision to re-optimize is made. In this situation, the optimization model considers the current location of the crew in the initial crew locations set $\mathcal{P}_{\mathcal{C}}$ and the location of the fault, which is the same location as the current location of the crew,  in the fault locations set $\mathcal{P}_{\mathcal{F}}$. Since the travel time $\Delta_{p, p^{\prime}}$ from the crew to the fault is zero, the optimization model can determine whether the crew should continue working on the current fault without delay or move to another location.}

\subsubsection{Action Times} Based on the movement paths of the crews, as defined in the previous constraints, and the timing for each repair or switching operation, we aim to determine the times that MS switching, fault repair, and zone patrol are completed. Generally, for a selected path, the time to perform a new action is calculated as the sum of the time taken to perform the previous action, the time for the crew to move to the new location, and the operation time for the new action:
\begin{subequations}
\begin{align}
& \tau_{p}^{c} \geq \tau_{p^{\prime}}^{c}+\Delta_{p^{\prime}, p}+T_{p}^{o p}+M\left(\beta_{p^{\prime}, p}-1\right) ; \quad \forall p, p^{\prime} \in \mathcal{P} \label{eq: actions times:1} \\
& \tau_{p}^{c} \geq \tau_{p^{\prime}}^{c}+M\left(\beta_{m}^{MSP}-1\right); \nonumber \\
& \qquad \qquad \qquad \forall m \equiv p, m \in \mathcal{M}, p \in \mathcal{P}_{\mathcal{M}}, p^{\prime} \in \mathcal{P}_{\mathcal{F}_{\mathcal{Q}}} \label{eq: actions times:2} \\
& \tau_{p}^{c} \geq B T\left(1-\beta_{p}^{V}\right) ; \quad \forall p \in \mathcal{P}_{\mathcal{M}^{\prime}} \label{eq: actions times:3} \\
& \tau_{p}^{c} \geq B T\left(1-\beta_{p}^{V}-\beta_{m}^{MSP}\right) ; \quad \nonumber \\
& \qquad \qquad \qquad \qquad \forall m \equiv p, m \in \mathcal{M}, p \in \mathcal{P}_{\mathcal{M}} \label{eq: actions times:4} \\
& \tau_{p^{\prime}}^{c} \geq \tau_{p}^{c} ; \quad \forall p \equiv p^{\prime}, p \in \mathcal{P}_{\mathcal{M}}, p^{\prime} \in \mathcal{P}_{\mathcal{M}^{\prime}}.\label{eq: actions times:5}
\end{align}
\label{eq: action times}
\end{subequations}
In these equations, $\mathcal{P}_{\mathcal{F}_{\mathcal{Q}}}$ denotes the set of locations of hypothetical faults, $\tau_{p}^{c}$ is the time when a crew completes its operations at location $p$, $\Delta_{p^{\prime}, p}$ is the travel time between locations $p^{\prime}$ and $p$, $T_{p}^{o p}$ is the operation time at location $p$, $M$ is a large constant, and $BT$ is a large out-of-scope time value.
Equation \eqref{eq: actions times:1} gives the earliest time $\tau_{p}^{c}$ at which a crew can finish its operations at the new location $p$, \eqref{eq: actions times:2} stipulates that the time for an MS opening through patrolling must be greater than the patrol time. If an MS is not switched for the first or second time, a large out-of-scope value is assigned to the switching time in \eqref{eq: actions times:3} and \eqref{eq: actions times:4}. \textcolor{black}{The second switching time for an MS exceeding $BT$ implies that the MS status remains unchanged, retaining the status after the first switching. Similarly, the first switching time for an MS greater than $BT$ indicates that the second switching time will also exceed $BT$ according to \eqref{eq: actions times:5}. This signifies that the MS status remains the same as its initial status.} As per \eqref{eq: actions times:5}, the second manual switching of an MS must occur after the first one.

\subsubsection{Network Reconfiguration Constraints}
The next set of constraints relates to the energization paths for each load or zone and governs whether parts of the network operate as isolated islands or remain connected to the upstream network:%
\begin{subequations}
\begin{align}
& \alpha_{z}^{\text {root }}=\sum_{b \in \mathcal{B}_{z}}\left\{\alpha_{b}^{\text {sub }}+\alpha_{b}^{DG}\right\} ; \quad \forall z \in \mathcal{Z} \label{eq: network reconfg:1} \\
& \alpha_{z}^{\text {root }}+\sum_{z^{\prime} \in \mathcal{Z}} \beta_{z^{\prime}, z}^{zz}=1 ; \quad \forall z \in \mathcal{Z} \label{eq: network reconfg:2} \\
& \beta_{z^{\prime}, z}^{zz}+\beta_{z, z^{\prime}}^{zz} \leq 1 ; \quad \forall z, z^{\prime} \in \mathcal{Z} \label{eq: network reconfg:3} \\
& \beta_{z^{\prime}, z}^{zz}+\beta_{z, z^{\prime}}^{zz}=\sum_{r \in \mathcal{R}_{z, z^{\prime}}} \beta_{r}^{RCS}+\sum_{m \in \mathcal{M}_{z, z^{\prime}}} \beta_{m}^{MSF} ; \forall z, z^{\prime} \in \mathcal{Z} \label{eq: network reconfg:4} \\
& \beta_{m}^{MSF}=\left\{\begin{array}{c}\beta_{m}^{MSI}\left(1-\beta_{p}^{V}-\beta_{m}^{MSP}+\beta_{p^{\prime}}^{V}\right) \\ +\left(1-\beta_{m}^{MSI}\right)\left(\beta_{p}^{V}-\beta_{p^{\prime}}^{V}\right)\end{array}\right\}; \nonumber \\
& \qquad \qquad \forall m \equiv p \equiv p^{\prime}, m \in \mathcal{M}, p \in \mathcal{P}_{\mathcal{M}}, p^{\prime} \in \mathcal{P}_{\mathcal{M}^{\prime}} \label{eq: network reconfg:5} \\
& \beta_{\ell}^{\text {line }}=\begin{cases}\beta_{r}^{RCS}; \, \ell=\mathcal{L}_{r}^{\text {RCS }} \\ \beta_{m}^{MSF};\, \ell=\mathcal{L}_{m}^{\text {MS }} \\ 1; \quad \text{otherwise} \end{cases} \forall \ell \in \mathcal{L}, m \in \mathcal{M}, r \in \mathcal{R},\label{eq: network reconfg:6}
\end{align}
\end{subequations}
where $\alpha_{z}^{\text {root}}$ is a binary variable indicating the power supply reference zone; $\alpha_{b}^{\text {sub }}$ is a binary value showing if bus $b$ is a substation; $\alpha_{b}^{DG}$ is a binary variable indicating if bus $b$ is hosting a master DG, i.e., a DG that remains separated from substations or other master DGs; and $\mathcal{B}_{z}$ is the set of buses in zone $z$. The term $\beta_{z^{\prime}, z}^{zz}$ is a binary variable indicating whether zone $z$ is energized by zone $z^{\prime}$; $\beta_{r}^{RCS}$ is a binary variable indicating the connection status of RCS $r$ in \eqref{eq: network reconfg:4}; $\mathcal{R}_{z, z^{\prime}}$ and $\mathcal{M}_{z, z^{\prime}}$ are the sets of all RCS and MSs between zone $z$ and $z^{\prime}$, respectively; and $\beta_{m}^{MSF}$ is a binary variable representing the final status of MS $m$. In \eqref{eq: network reconfg:5}, $\beta_{m}^{MSI}$ represents the initial status of MS $m$. In \eqref{eq: network reconfg:6}, $\beta_{\ell}^{\text {line }}$ is the final line connection status, and $\mathcal{L}_{m}^{\text {MS }}$ and $\mathcal{L}_{r}^{\text {RCS }}$ are the lines switchable by MS $m$ and RCS~$r$, respectively.

\textcolor{black}{Each zone containing a substation is a reference zone. Other zones having DGs but not having a substation can also serve as reference zones and initiate energization paths. For such zones, a specific DG must be designated as the ``master DG,'' denoted by the optimization model ($\alpha^{DG}_b=1$). Therefore, t}he reference zone includes a substation bus or a master DG \eqref{eq: network reconfg:1}, so it is not energized through another zone. This statement is reflected in \eqref{eq: network reconfg:2} which indicates that each zone is either a reference zone or is energized by another zone. This condition also implies maintaining the radial structure of the network. As described in \eqref{eq: network reconfg:3}, for a pair of zones in the network, only one zone can energize the other (parent/child relation). In a complex network structure, it is possible for two zones to be connected via multiple switches. If one zone energizes another zone, only one switch (RCS or MS) between the two zones must be in the connected state \eqref{eq: network reconfg:4}. The MS final status is calculated based on its initial status and switching actions in \eqref{eq: network reconfg:5}. The final line connection status is calculated based on the final MS or RCS status \eqref{eq: network reconfg:6}.

\subsubsection{Zone Restoration Times} So far, the constraints related to the energization path of the network zones, switching, and repair times have been introduced. Knowing these values, the outage duration (energization time) of different zones is calculated. The parent must be energized before the child for each for each pair of connected zones:
\begin{equation}
\label{eq: Zones restoration time:1}
T_{z}^{\text {out }} \geq T_{z^{\prime}}^{\text {out }}-M\left(1-\beta_{z^{\prime}, z}^{zz}\right) ; \quad \forall z, z^{\prime} \in \mathcal{Z},
\end{equation}
where $T_{z}^{\text {out }}$ is the outage duration of zone $z$.
If an MS isolates two zones, the zones on each side of the switch cannot have a restoration time smaller than the switching time. Before the MS is opened, these two zones are connected and they thus cannot be restored due to the lack of fault isolation or violations of technical constraints:
\begin{align}
\label{eq: Zones restoration time:2}
& T_{z}^{\text {out }} \geq \tau_{p}^{c}-M\left(1-\beta_{p}^{V}-\beta_{m}^{MSP}\right); \nonumber \\ & 
 \quad  \forall z, z^{\prime} \in \mathcal{Z}, m \in \mathcal{M}_{z, z^{\prime}}, p \equiv m, p \in \mathcal{P}_{\mathcal{M}}, \beta_{m}^{MSI}=1,
\end{align}
where $\mathcal{M}_{z, z^{\prime}}$ is the set of all MSs connecting zones $z$ and $z^{\prime}$ and $\beta_{m}^{MSI}=1$ indicates that MS $m$ is initially closed. 
If an MS is finally closed after a second switching, one of its connected zones will be the parent and the other will be the child. In this case, according to the description of the load restoration process, first, the switch is opened in order to separate the two zones and energize the parent zone, and then it is closed again in order to restore the child zone. Therefore, only the child zone will have a restoration time greater than the second switching time:
\begin{align}
\label{eq: Zones restoration time:3}
& T_{z}^{\text {out }} \geq \tau_{p}^{c}-M\left(2-\beta_{z^{\prime}, z}^{zz}-\beta_{p}^{V}\right); \nonumber \\
& \quad \forall z, z^{\prime} \in \mathcal{Z}, m \in \mathcal{M}_{z, z^{\prime}}^{\prime}, p \equiv m, p \in \mathcal{P}_{\mathcal{M}^{\prime}}, \beta_{m^{\prime}}^{MSI}=1,
\end{align}
where $\mathcal{M}_{z, z^{\prime}}^{\prime}$ represents the set of MSs connecting $z$ and $z^{\prime}$ for second switching actions.
For a child zone restored by closing a normally open MS, the zone restoration time will be greater than the manual switching time: 
\begin{align}
\label{eq: Zones restoration time:4}
 & T_{z}^{\text {out }} \geq  \tau_{p}^{c}-M\left(2-\beta_{z^{\prime}, z}^{zz}-\beta_{p}^{V}\right); \nonumber \\
& \quad \forall z, z^{\prime} \in \mathcal{Z}, m \in \mathcal{M}_{z, z^{\prime}}, p \equiv m, p \in \mathcal{P}_{\mathcal{M}}, \beta_{m}^{MSI}=0.
\end{align}
If a normally closed MS remains closed, it will surely energize one of the two zones on its two sides. In this situation, the parent zone cannot be energized before the child zone because these zones are connected during the entire procedure:
\begin{align}
\label{eq: Zones restoration time:5}
& T_{z}^{\text {out }} \geq T_{z^{\prime}}^{\text {out }}-M\left(1-\beta_{z, z^{\prime}}^{zz}+\beta_{p}^{V}+\beta_{m}^{MSP}\right); \nonumber \\
& \forall z, z^{\prime} \in \mathcal{Z}, m \in \mathcal{M}_{z, z^{\prime}}, p \equiv m, p \in \mathcal{P}_{\mathcal{M}}, \beta_{m}^{MSI}=1.
\end{align}
A zone cannot be energized until all related faults have been repaired. Therefore, the time to restore a zone must be longer than the time to repair all the faults in that zone:
\begin{equation}
\label{eq: Zones restoration time:6}
T_{z}^{\text {out }} \geq \tau_{p}^{c} ; \quad \forall z \in \mathcal{Z}, f \in \mathcal{F}_{z}, p \equiv f, p \in \mathcal{P}_{\mathcal{F}},
\end{equation}
where $\mathcal{F}_{z}$ is the set of all faults in zone $z$ and $\mathcal{P}_{\mathcal{F}}$ is the set of all locations $p$ with a fault.
If an MS were closed before all faults are repaired in the child zone, the parent zone would be subject to the repair time of the offspring zone. Therefore, it is preferred that the MS does not have a closing time earlier than the restoration time of the child zone:
%
%
{\small
\begin{align}
\label{eq: Zones restoration time:7}
&\tau_{p}^{c}\left(1-\beta_{m}^{MSI}\right)+\tau_{p^{\prime}}^{c}\left(\beta_{m}^{MSI}\right) \geq T_{z}^{\text {out }}-M\left(1-\beta_{z^{\prime}, z}^{zz}\right) ;\nonumber \\
&\forall z, z^{\prime} \in \mathcal{Z}, m \in \mathcal{M}_{z, z^{\prime}},m\equiv p \equiv p^{\prime},p \in \mathcal{P}_{\mathcal{M}},p^{\prime} \in \mathcal{P}_{\mathcal{M}^{\prime}}.
\end{align}
}
In essence, constraints~\eqref{eq: Zones restoration time:1}--\eqref{eq: Zones restoration time:7} define the interactions between zones, switches, and repair activities during restoration.

\subsubsection{Power Flow Expression}
Here, we discuss power flow expressions, which consist of multi-time-step conventional power flow (PF) and a time-step free conservative PF in accordance with the proposed routing framework. 

The conventional PF model is discussed first. Consider two generic zones able to be connected by a switch that possibly have their own active$\backslash$reactive power injections. When a zone $z$ energizes another zone $z'$ after its own energization, the power flow and voltage conditions may change in zone $z$. Therefore, for a network with $n$ zones, $n$ sets of power flow equations are required to guarantee safe voltage and line flow values. 
During the restoration process, each step involves the energization of one zone, and one set of power flow constraints is added at each step. These constraints are:
\begin{subequations}
\label{power_flow}
\begin{align}
& \zeta_{z, z^{\prime}} \geq\left(T_{z^{\prime}}^{\text {out }}-T_{z}^{\text {out }}\right) / T^{\text {max }} ; \quad \forall z, z^{\prime} \in \mathcal{Z} \label{eq: power flow:1} \\
& \sum_{z \in \mathcal{Z}} \beta_{z, t}^{zt}=t ; \quad \forall t \in \mathcal{T} \label{eq: power flow:2} \\
& \beta_{z, t}^{zt} \geq \beta_{z, t-1}^{zt} ; \quad \forall t \in \mathcal{T}, z \in \mathcal{Z} \label{eq: power flow:3} \\
& \sum_{t \in \mathcal{T}}\left(\beta_{z, t}^{zt}-\beta_{z^{\prime}, t}^{zt}\right) \geq 1-\left(1-\zeta_{z, z^{\prime}}\right) M ; \quad \forall z, z^{\prime} \in \mathcal{Z} \label{eq: power flow:4} \\
& \sum_{t \in \mathcal{T}}\left(\beta_{z, t}^{zt}-\beta_{z^{\prime}, t}^{zt}\right) \geq 1-\left(1-\beta_{z, z^{\prime}}^{zz}\right) M ; \;\;\, \forall z, z^{\prime} \in \mathcal{Z}, \label{eq: power flow:5}
\end{align}
\end{subequations}
where $\zeta_{z, z^{\prime}}$ is a binary variable indicating earlier energization time for zone $z$ than $z^{\prime}$ and $T^{\text {max }}$ is the maximum possible outage time.
The binary variable $\beta_{z, t}^{zt}$ in \eqref{eq: power flow:2} and \eqref{eq: power flow:3} tracks the energization status of each zone at each time step, where $t$ represents a time step and $\mathcal{T}$ is the set of all time steps. In each time step, one zone is energized \eqref{eq: power flow:2} and remains at that state for the rest of the process \eqref{eq: power flow:3}.
For any pair of zones  $z$ and $z^{\prime}$, if $z$ is energized earlier ($\zeta_{z, z^{\prime}}=1$ in \eqref{eq: power flow:4}) or is the parent zone ($\beta_{z, z^{\prime}}^{zz}=1$ in \eqref{eq: power flow:5}), then it has been in the network for more time steps. Otherwise, the constraints are relaxed by a large margin of $M$. These constraints ensure that power is transferred in the correct order following the energization paths.

A set of power flow equations consists of voltage drop equation \eqref{eq: power flow cont:1} (see~\cite{low2014pscc} for details on the model we use in this paper), power balance \eqref{eq: power flow cont:2}, power source limitations \eqref{eq: power flow cont:3}, and voltage and line flow limits (e.g., see \cite{ahmadi2015}):%
\begin{subequations}
\label{eq: power flow cont}
\begin{align}
& \pm \color{black}F\color{black}\left(U_{b, t}, \varphi_{\ell, t}\right) \leq M\left(1-\beta_{\ell}^{\text {line }}\right) \label{eq: power flow cont:1} \\
& \sum_{\ell \sim b} \varphi_{\ell, t}+\beta_{z, t}^{zt} D_{b}-G_{b, t}=0 ;\space\forall b \in \mathcal{B}_{z}, z \in \mathcal{Z}, t \in \mathcal{T} \label{eq: power flow cont:2} \\
& \beta_{z, t}^{zt} G_{b}^{\text {min }} \leq G_{b, t} \leq \beta_{z, t}^{zt} G_{b}^{\text {max }} ; \space\forall b \in \mathcal{B}_{z}, z \in \mathcal{Z}, t \in \mathcal{T}, \label{eq: power flow cont:3}\\
& \color{black}U^{min}\leq U_{b,t}\leq U^{max},\quad  \left|\varphi_{\ell, t}\right|\leq\varphi^{max}, \label{eq: power flow cont:4}
\end{align}
\end{subequations}
where $\color{black}F\color{black}\left(U_{b, t}, \varphi_{\ell, t}\right)$ in \eqref{eq: power flow cont:1} represents the voltage drop as a function of the voltage magnitude at bus $b$ at time $t$, $U_{b, t}$, and the flow of line $\ell$ at time $t$, $\varphi_{\ell, t}$.
In \eqref{eq: power flow cont:2}, the summation term represents the total power flow export from bus $b$ to all lines connected to it, denoted as $\ell \sim b$, $D_{b}$ is the demand at bus $b$, and $G_{b, t}$ is the generation at bus $b$ at time $t$. Variables $\varphi$, $D$, and $G$ concisely represent both active and reactive powers.
In \eqref{eq: power flow cont:3}, $G_{b}^{\text {min }}$ and $G_{b}^{\text {max }}$ are the minimum and maximum generation at bus~$b$, respectively. \textcolor{black}{Maintaining voltage levels and line flows within statutory ranges are ensured by~\eqref{eq: power flow cont:4} (see \cite{ahmadi2015} for a linear approximation of the flow limits).}

The concept of conservative PF is introduced to reduce the computational burden of solving multiple sets of power flow equations. \textcolor{black}{In the final network configuration, two distinct loading conditions with different sets of variables, namely passive loading and active loading, collectively establish} upper and lower bounds for nodes' voltage levels and an upper bound for lines' power flows throughout every steps of the restoration process. To acquire these bounds, in this paper, we assume that as the result of a good switch placement strategy in a previous planning stage, our network loading in electrical zones is nearly three-phase balanced \cite{chen2018}, the condition at which voltage levels and line loading conditions monotonically change along the feeder \cite{kersting2012distribution}.

\textit{\textbf{Passive loading}}: The purpose of this loading condition is to determine a lower bound for voltage levels, $L B\{U\}$, in all restoration steps. This condition is referred to as ``passive loading.'' Since DGs and capacitors can supply power up to their designated zone’s and downstream zones’ aggregate demand while respecting their own generation upper limits. However, in the passive loading condition, active and reactive injections from downstream zones to upstream zones are prohibited. Furthermore, the DGs’ lower power generation limits are also relaxed to accommodate cases where the lower generation limits are above the total power consumption within their designated zone’s and downstream zones’. Consequently, each zone treats all of its downstream zones as a collective passive load as shown in Fig.~\ref{fig:  Passive loading}. Passive loading lowers voltage levels as new zones are restored. For the passive loading condition, the power flow constraints are:
\begin{figure}[t]
    \centering
    \includegraphics [width=2.8in]{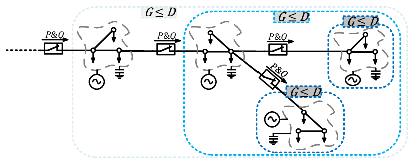}
    \vspace{-0.5em}
    \caption{Passive loading condition}
    \label{fig:  Passive loading}
    \vspace{-1em}
\end{figure}
\begin{subequations}
\label{eq: conservative PF}
\begin{align}
& \pm \color{black}F\color{black}\left(\underline{U}_{b}, \underline{\varphi}_{\ell}\right) \leq M\left(1-\beta_{\ell}^{\text {line }}\right) \label{eq: Passive loading:1} \\
& \sum_{\ell \sim b} \underline{\varphi}_{\ell}+D_{b}-\underline{G}_{b}=0 ; \quad \forall b \in \mathcal{B} \label{eq: Passive loading:2} \\
& \underline{G}_{b} \leq G_{b}^{\max } ; \quad \forall b \in \mathcal{B} \label{eq: Passive loading:3} \\
& \underline{\varphi}_{\ell} \geq\left(\zeta_{z, z^{\prime}}-1\right) M ; \quad \forall z, z^{\prime} \in \mathcal{F}, \label{eq: Passive loading:4}\\
& \color{black}U^{min}\leq \underline{U}_{b}\leq U^{max},\quad  \left|\underline{\varphi}_{\ell}\right|\leq\varphi^{max}, \label{eq: Passive loading:5}
\end{align}%
\end{subequations}%
where $\underline{U}_{b}$ is voltage magnitude at bus $b$,  $\underline{\varphi}_{\ell}$ is the flow on line $\ell$, and $\underline{G}_{b}$ is the generation at bus $b$, respectively, all in the passive loading condition.
If there is a time difference between energization of $z$ and $z^{\prime}$, then $z^{\prime}$ is added as a passive load $\left(\underline{\varphi}_{\ell} \geq 0\right)$ to $z$ as shown in \eqref{eq: Passive loading:4}. This constraint reduces nodes' voltage levels monotonically by adding a new zone. In \eqref{eq: Passive loading:3}, lower bounds on the power generation are relaxed since $\sum_{b \in z^{\prime}} D_{b} \leq \sum_{b \in z^{\prime}} G_{b}^{\min }$ forces $T_{z}^{\text {out }}=T_{Z^{\prime}}^{\text {out }}$ such that $z^{\prime}$ can send the extra generated power to $z$.

\begin{figure}[t]
    \centering
    \includegraphics [width=2.8in]{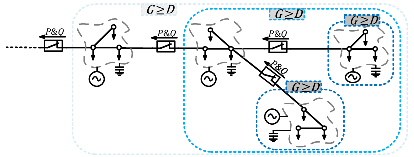}
    \vspace{-0.5em}
    \caption{Active loading condition}
    \label{fig:  Active loading}
    \vspace{-1em}
\end{figure}

\textit{\textbf{Active loading}}: In the active loading condition, new zones are added as active loads $(\overline{\varphi}_{\ell} \leq 0)$, leading to higher voltage levels. 
This paradigm is termed ``active loading'' because a specific zone accommodates DGs within its domain along with power injections from downstream zones to meet the entire demand within the zone, potentially allowing for power export to the upstream zone.
%
However, the outbound power transmission from a zone to its downstream counterparts remains prohibited. To be able to generate that much power, the upper limit of the DGs' power generation is relaxed, permitting them to produce power beyond their rated capacities. This approach additionally considers the presence of a DG at every node. As a result, each zone treats its entire downstream network as an active load as shown in Fig.~\ref{fig:  Active loading}. As new zones are restored, active loading increases voltage levels. Thus, if the voltage levels for the final configuration are within the acceptable range, the voltage levels of all preceding configurations also satisfy the voltage limits. 
For the active loading condition, the power flow constraints are:
\begin{subequations}\label{eq:  active loading}
\begin{align}
& \pm \color{black}F\color{black}\left(\overline{U}_{b}, \overline{\varphi}_{\ell}\right) \leq M\left(1-\beta_{\ell}^{\text {line }}\right) \label{eq:  active loading:1} \\
& \sum_{\ell \sim b} \overline{\varphi}_{\ell}+D_{b}-\overline{G}_{b}=0 ; \quad \forall b \in \mathcal{B} \label{eq:  active loading:2} \\
& G_{b}^{\text {min }} \leq \overline{G}_{b} ; \quad \forall b \in \mathcal{B} \label{eq:  active loading:3} \\
& \underline{G}_{b} \leq \overline{G}_{b} ; \quad \forall b \in \mathcal{B} \label{eq:  active loading:4} \\
& \overline{\varphi}_{\ell} \leq\left(1-\zeta_{z, z^{\prime}}\right) M ; \quad \forall z, z^{\prime} \in \mathcal{Z}, \label{eq:  active loading:5}\\
& \color{black}U^{min}\leq \overline{U}_{b}\leq U^{max},\quad  \left|\overline{\varphi}_{\ell}\right|\leq\varphi^{max}, \label{eq: active loading:6}
\end{align}
\end{subequations}
where $\overline{U}_{b}$ is the voltage level at bus $b$, $\overline{\varphi}_{\ell}$ is the flow on line $\ell$, and $\overline{G}_{b}$ is the generation at bus $b$, respectively, for the active loading condition. As shown in Section~\ref{base-case-study}, our numerical results validate the accuracy of the power flow linearization from \cite{low2014pscc} for our formulation, with voltage magnitudes within 0.0058 per unit of the nonlinear AC power flow model.
The appendix 
\ifarxiv
\else
in \cite{jalilian2023cooptimization}
\fi
provides derivations showing how the passive and active loading conditions result in upper and lower bounds on the voltages and upper bounds on the line flows with respect to the power flow approximation’s outputs.

\begin{figure}[t]
    \centering
    \includegraphics [width=2.8in]{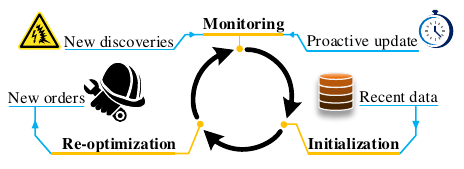}
    \vspace{-0.5em}
    \caption{\textcolor{black}{Overall workflow summary}}
    \label{fig:  overall workflow}
    \vspace{-1.5em}
\end{figure}

\textcolor{black}{The full optimization problem can be described either in a multi-time-step or time-step-free approach.  The multi-time-step formulation is:
\begin{subequations}\label{eq:  multi-time-step}
\begin{align}
& \min: \quad (2) \\
& \text{s.t.} \quad \text{(3)\,--\,(14)}
\end{align}
\end{subequations}
The time-step-free formulation is:
\begin{subequations}\label{eq:  time-step-free}
\begin{align}
& \min: \quad (2) \\
& \text{s.t.} \quad \text{(3)\,--\,(13a), (15), (16)}
\end{align}
\end{subequations}}
\vspace{-2em}
\textcolor{black}{\subsection{Overall Workflow Summary}}\label{Overall_Workflow_Summary
}
\textcolor{black}{To summarize, our proposed method provides a holistic and dynamic framework for optimizing the decision-making process in fault management using the following three phases, as shown in Fig.~\ref{fig:  overall workflow}.}\\
\textcolor{black}{\textit{\textbf{Initialization Phase:}} The first phase initializes the most recent values for the model including information regarding the power system network, repair crews, repair times, fault probabilities, fault locations, and travel times between locations in the network. This includes assigning hypothetical faults to each unpatrolled zone with expected repair times as defined in~\eqref{eq: time}.}\\
\textcolor{black}{\textit{\textbf{Re-optimization phase:}} Using the latest data, the second phase solves the optimization problem \eqref{eq:  multi-time-step} or \eqref{eq:  time-step-free}. This process updates previous decisions while incorporating the latest real-time information to enhance the system's adaptability and effectiveness in restoring service. Accordingly, repair crews are informed with new orders. }\\
\textcolor{black}{\textit{\textbf{Monitoring phase:}} The third phase continuously monitors real-time information from the repair crews to facilitate dynamic decision-making. The monitoring phase incorporates two schemes: responsive re-optimization and proactive re-optimization. Under responsive re-optimization, our method responds to specific events referred to as ``new discoveries.'' These events include completing the patrol of a zone as well as identifying the location and assessing the repair time for a new fault. To prevent premature responses, we introduce a minimum update time (e.g., 10 minutes) such that our method does not respond sooner than the predefined minimum update time even if a new discovery event occurs. Our numerical studies in Section~\ref{subsec:Decision_Update_Frequency} indicate that this minimum update time can speed up the overall restoration process as faster responses to every new discovery do not always yield faster load restoration. The proactive re-optimization scheme defines a maximum update time by triggering a decision update if no new discovery occurs within a predefined interval (e.g., 30 minutes). This proactive approach ensures that our system remains adaptive even in the absence of new discoveries, contributing to the efficiency and responsiveness of our proposed method. This proactive adaptation is necessary due to other continuously changing parameters, such as travel times and the extent of unpatrolled zones, as crews continuously patrol the network. Upon triggering a re-optimization by either the responsive or proactive schemes, the method returns to the initialization phase and then to the re-optimization phase, as shown in Fig.~\ref{fig:  overall workflow}.
}

\vspace{0.5em}
\section{Numerical Results}\label{sec: Numerical Results}
\begin{figure*}[h!]
\centering
\captionsetup[subfloat]{labelfont=scriptsize,textfont=scriptsize}
\vspace{-2.25em}
\subfloat[1st step]{%
  \includegraphics[width=0.32\textwidth]{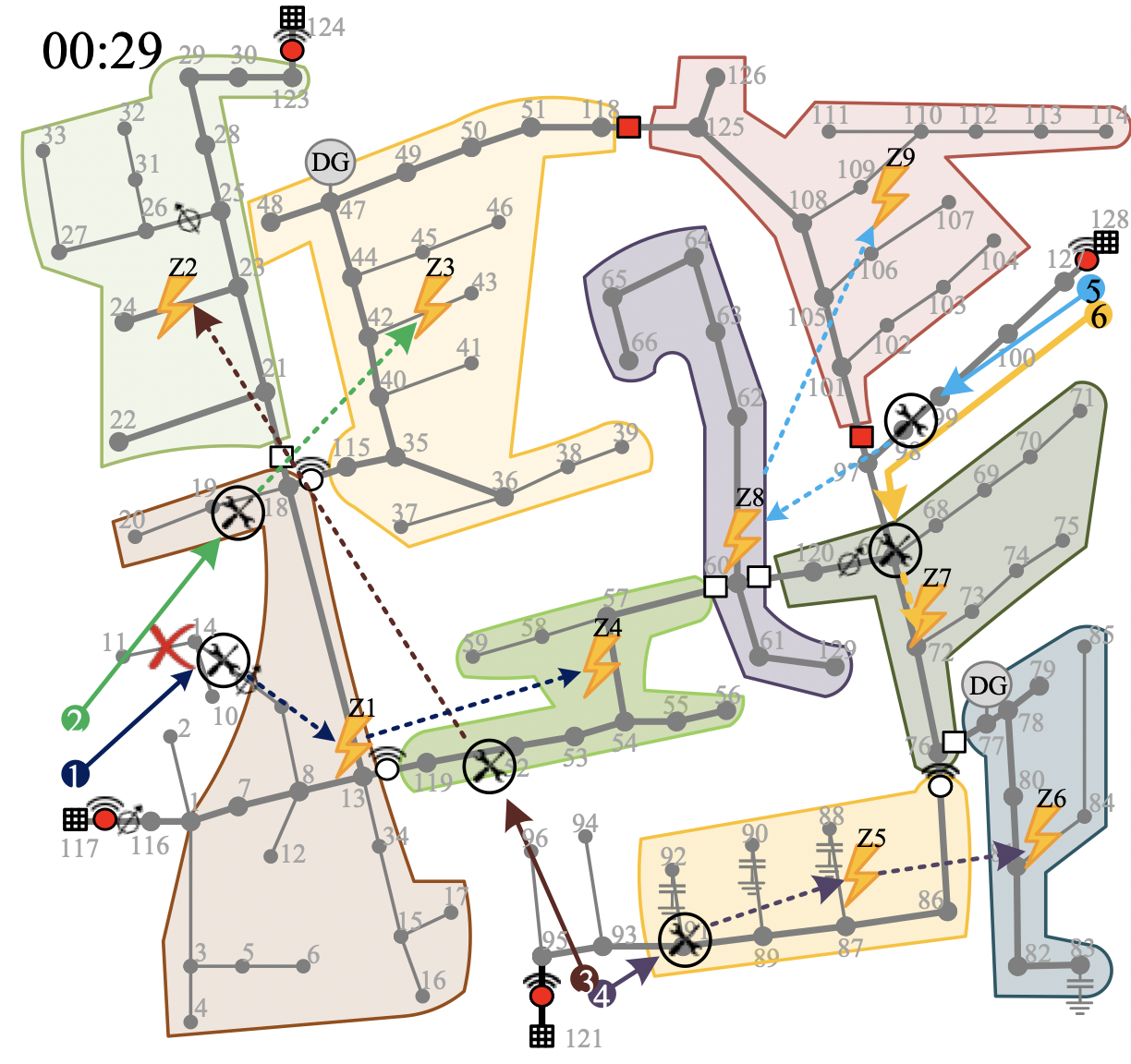}%
  \label{fig: sub1}%
}%
\hfill
\subfloat[2nd step]{%
  \includegraphics[width=0.3\textwidth]{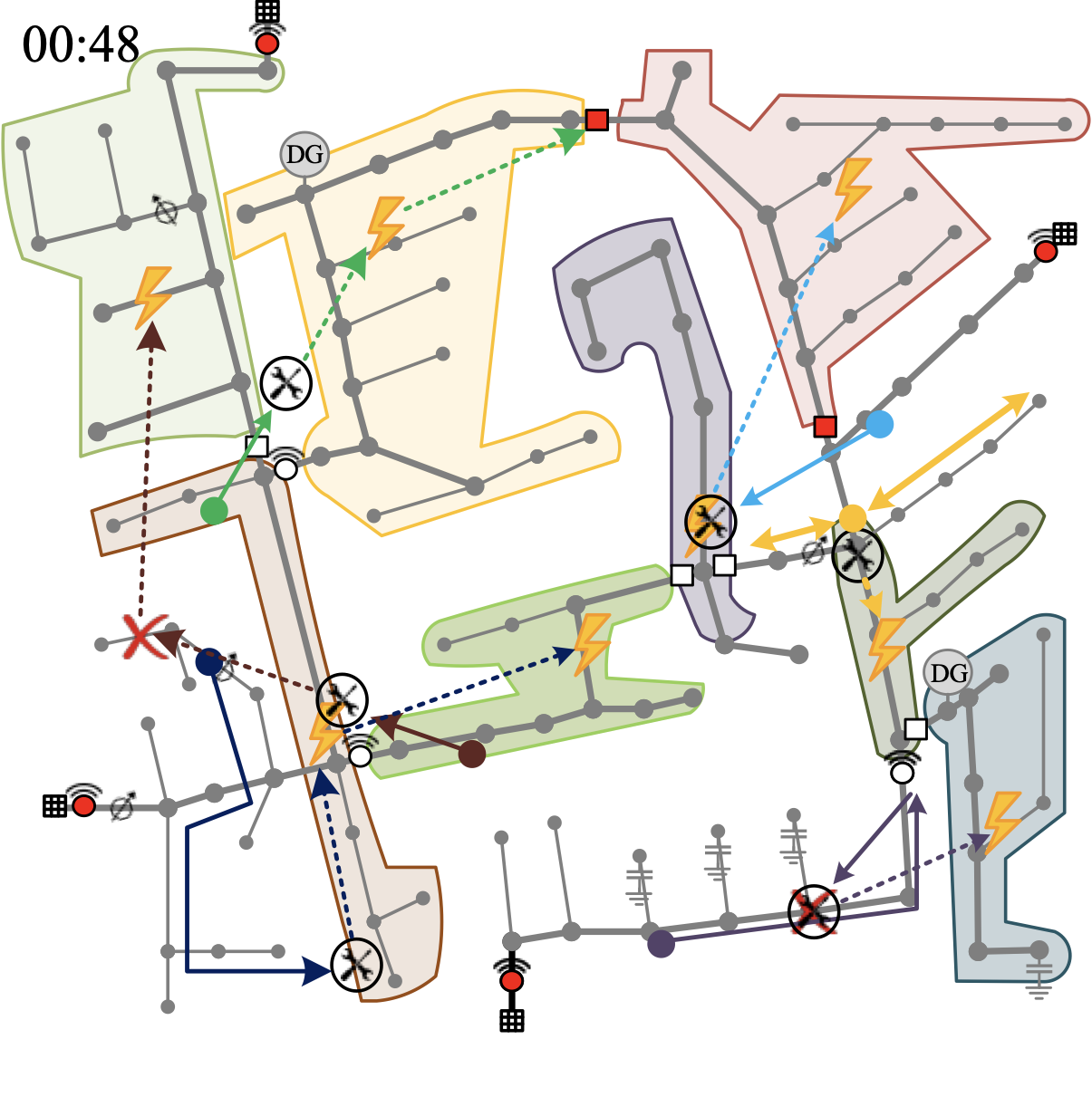}%
  \label{fig: sub2}%
}%
\hfill
\subfloat[5th step]{%
  \includegraphics[width=0.32\textwidth]{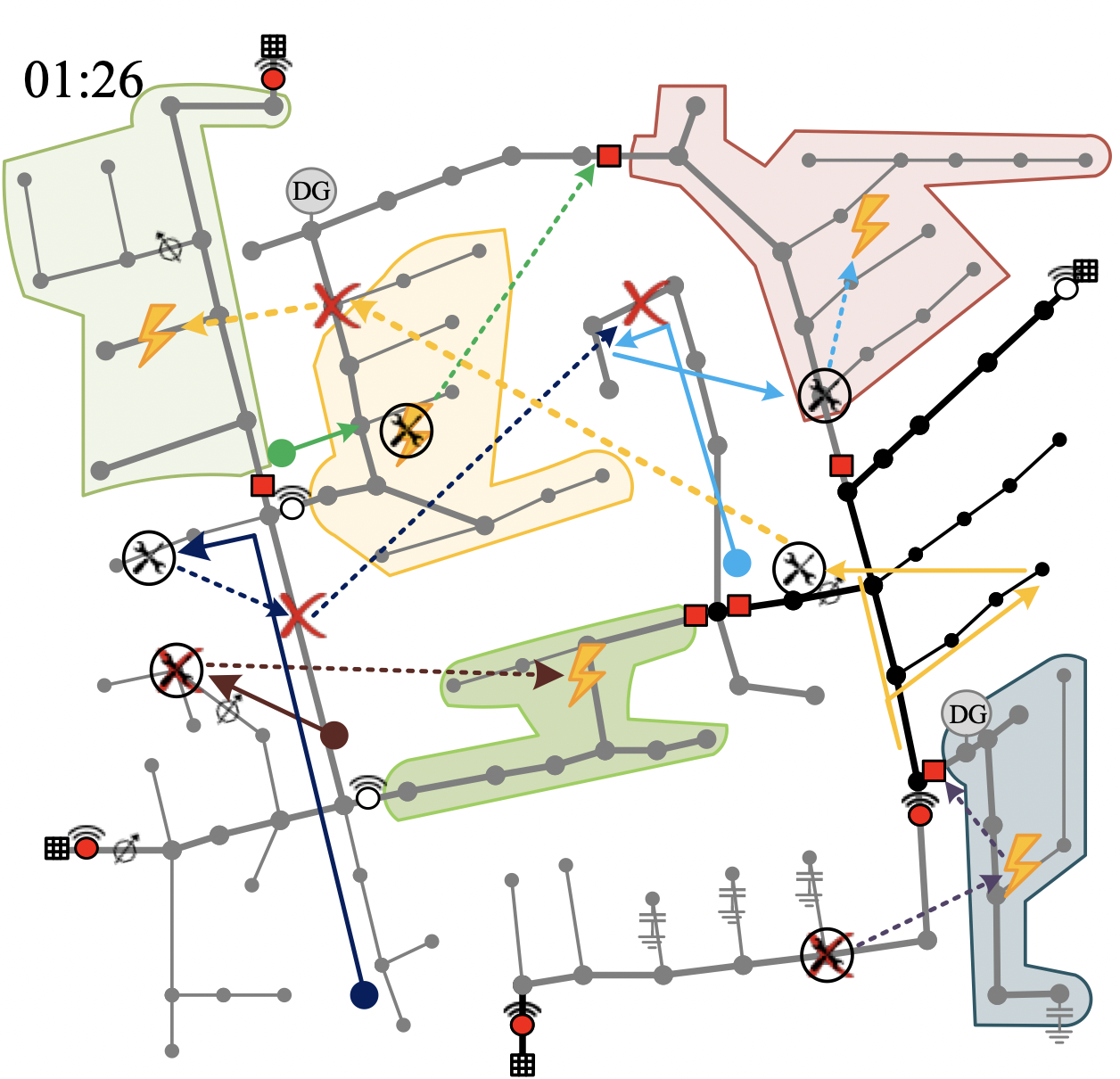}%
  \label{fig: sub3}%
}

\medskip
\vspace{-1.5em}
\subfloat[8th step]{%
  \includegraphics[width=0.3\textwidth]{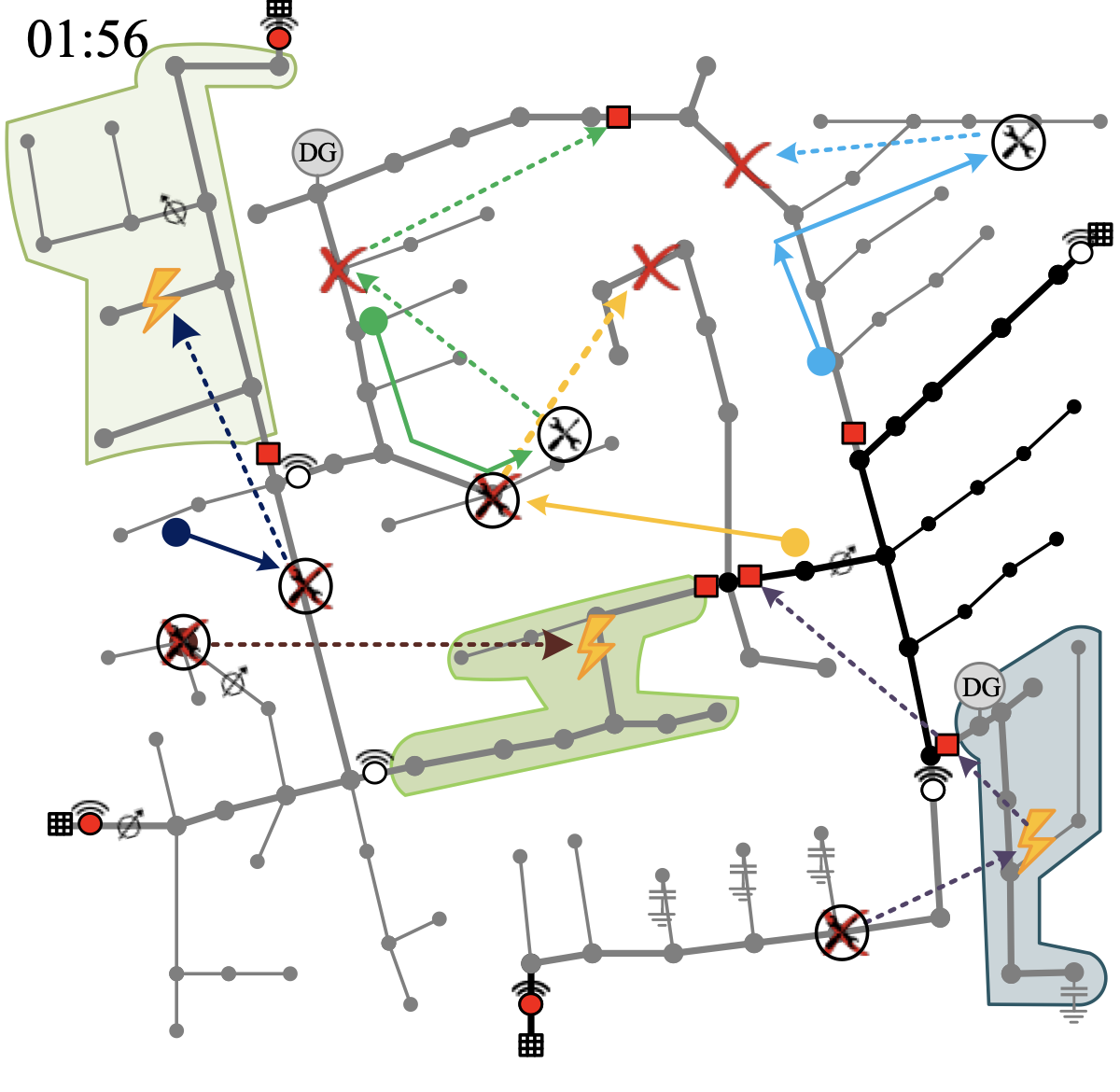}%
  \label{fig: sub4}%
}%
\hfill
\subfloat[10th step]{%
  \includegraphics[width=0.3\textwidth]{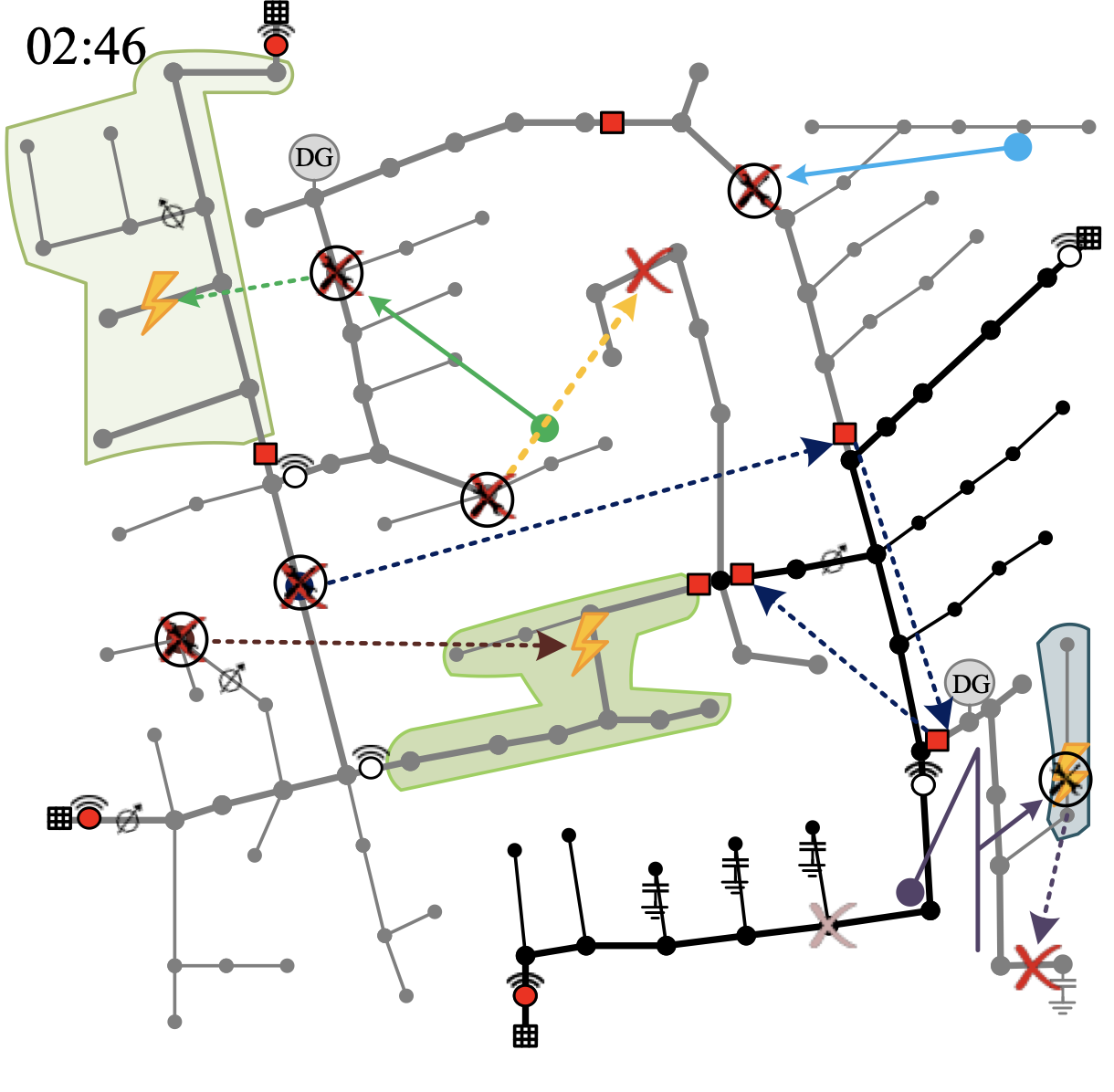}%
  \label{fig: sub5}%
}%
\hfill
\subfloat[16th step]{%
  \includegraphics[width=0.3\textwidth]{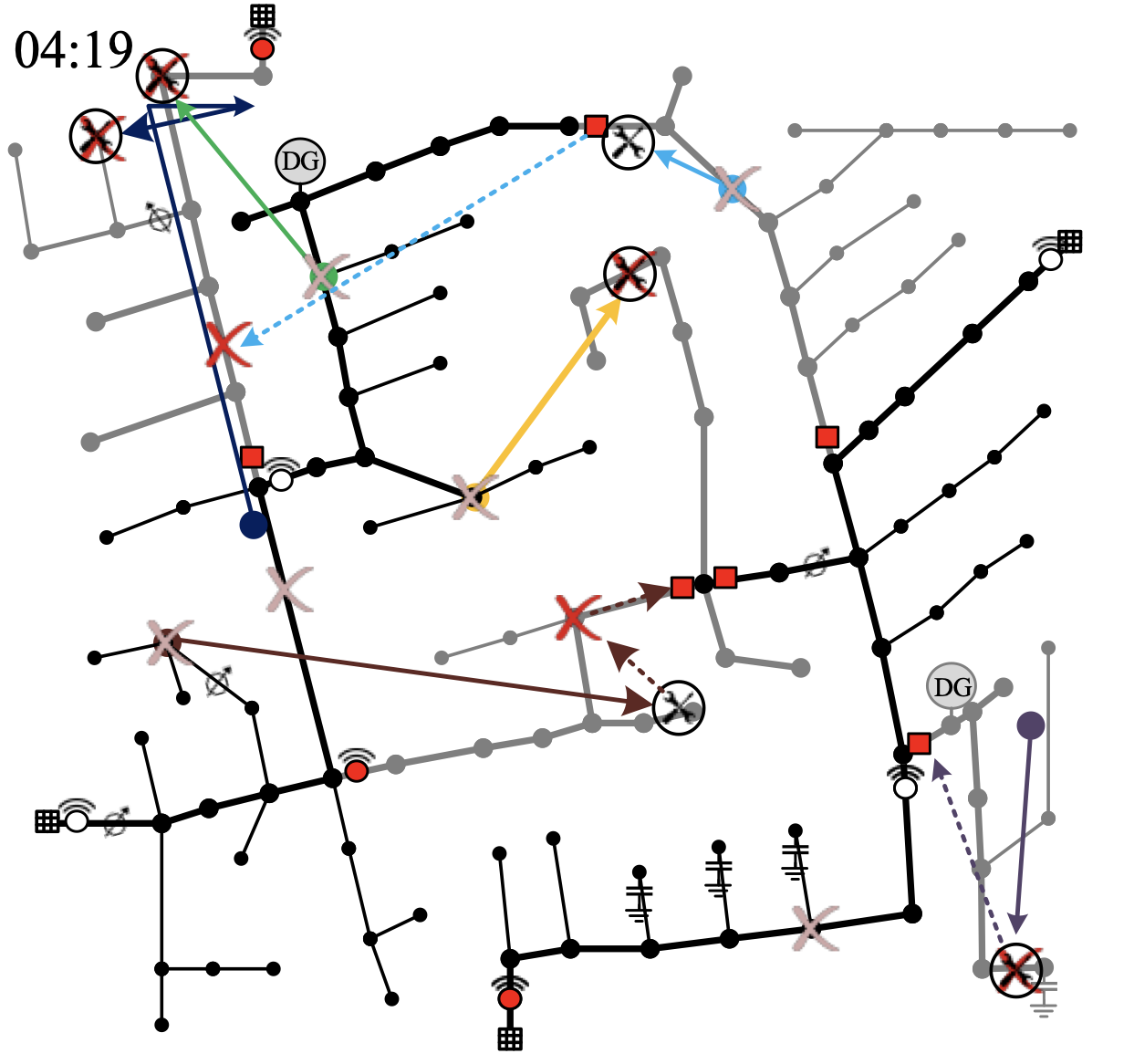}%
  \label{fig: sub6}%
}
\vspace{-1em}
\subfloat{%
  \includegraphics[width=0.9\textwidth]{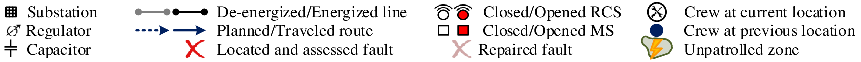}%
  \label{fig: sub7}%
}
\vspace{-0.5em}
\caption{Fault restoration process in IEEE 123-node test feeder; see \cite{Taheri2023} for an animation of the restoration procedure}
\label{fig:  fault management}
\end{figure*}

This section empirically evaluates the proposed model using modified \texttt{IEEE 123-node} and \texttt{IEEE 8500-node} \cite{8500-node} networks. The simulations have been designed to validate the model's efficiency and scalability.
The 123-node network shown in Fig.~\ref{fig:  fault management} includes $6$ MSs and $7$ RCSs, dividing the network into $13$ distinct zones. For the purposes of these studies, we assume that the operation time for MSs is $5$ minutes, while the operation time for RCSs is negligible. The parameters associated with the $2$ DGs are presented in Table \ref{tab: DG Parameters}.
In our simulated scenarios, system outages are triggered by $12$ faults, the locations and estimated repair times of which are detailed in Table \ref{tab:  Fault Parameters}. We assume that these parameters are unknown immediately post-event and are revealed progressively during the feeder patrolling process.
\begin{table}[h!]
\vspace{-1.5em}
\centering
\smaller
\caption{DG Parameters}
\vspace{-0.9em}
\label{tab: DG Parameters}
\renewcommand{\arraystretch}{1.5}
\begin{tabular}{c|ccc}
\hline 
\textbf{Name} & \textbf{Location} & $\overline{\boldsymbol{P}}^{\boldsymbol{DG}} / \underline{\boldsymbol{P}}^{\boldsymbol{DG}}$ & $\overline{\boldsymbol{Q}}^{\boldsymbol{DG}}  /\underline{\boldsymbol{Q}}^{\boldsymbol{DG}}$ \\
\hline 
DG1 & Bus 47 & $200/ 20$ kW & $\pm 140$ kW \\
\hline 
DG2 & Bus 77 & $300/ 30$ kW & $\pm 210$ kW\\
\hline
\end{tabular}
\vspace{-1em}
\end{table}
\begin{table*}[h!]
\centering
\smaller
\caption{Repair Time for Different Fault Locations (minutes)}
\vspace{-1em}
\label{tab: Fault Parameters}
\renewcommand{\arraystretch}{1}
\setlength{\tabcolsep}{3.03pt} 

\begin{tabularx}{\textwidth}{X*{12}{>{\centering\arraybackslash}X}}
\toprule
& \multicolumn{12}{c}{\textbf{Location}} \\
\cmidrule(lr){2-13}
& Line 14 & Line 33 & Bus 44 & Line 13 & Bus 87 & Line 84 & Bus 57 & Bus 108 & Line 64 & Line 22 & Bus 29 & Bus 36 \\
\midrule
\textbf{Time}& 120 & 100 & 75 & 80 & 70 & 74 & 60 & 110 & 160 & 59 & 75 & 90 \\
\bottomrule
\end{tabularx}
\end{table*}

We have also assumed the availability of $6$ crew teams for field operations, with patrol zones identical to the electrical zones for the sake of clarity. The cost of damage to customers is selected randomly from $\$15$ to $\$45$ per kWh, and the travel cost for crews is set at $\$0.60$ per hour of driving time. Travel times are calculated based on the straight distance between any pair of locations in the routing problem.

\vspace{-1em}
\subsection{Base Case Evaluation}\label{base-case-study}
In the aftermath of an extreme event, the breakers at the substations activate, and all load points experience an interruption. After executing our proposed optimization model, Fig.~\ref{fig:  fault management} illustrates the sequence of actions needed to restore service to the affected load points. The total restoration process spans $21$ optimization steps and a duration of $6$ hours and $36$ minutes, during which all load points are re-energized.
The timing of decision updates is contingent upon the completion of zone patrols or the detection and assessment of a fault; otherwise, the timing defaults to a set value (in this case, $30$ minutes). The update time never falls below a minimum length (in this case, $10$ minutes).
Fig.~\ref{fig:  fault management} presents the final moments of six selected steps from a total of $21$ steps. Solid arrows connect each crew's previous location (the initial location in the time step) to its current location (the final location in the time step), illustrating their path of travel. Dashed arrows show the crews' planned routes based on the latest set of decisions, which could be altered by subsequent decisions.
For example, at the start ($t=0$), crew $1$ is scheduled to patrol zones Z1 and Z4, and crew $3$ is designated for zone Z2 as shown in Fig.~\ref{fig: sub1}. However, at $t=29$, a fault is discovered in zone Z1 by crew $1$. Consequently, the routes are updated as shown in Fig.~\ref{fig: sub2}, with crew $3$ being reassigned to repair the fault before patrolling Z2.
As shown in Fig.~\ref{fig: sub3}, zone Z7 is isolated through during-patrol MS operation and energized since no damage is detected in that area. It is also worth noting that some crews are already engaging in repair and restoration operations while some zones are still pending patrol.

\begin{figure}[h]
\vspace{-1em}
\centering
\includegraphics[width=3in]{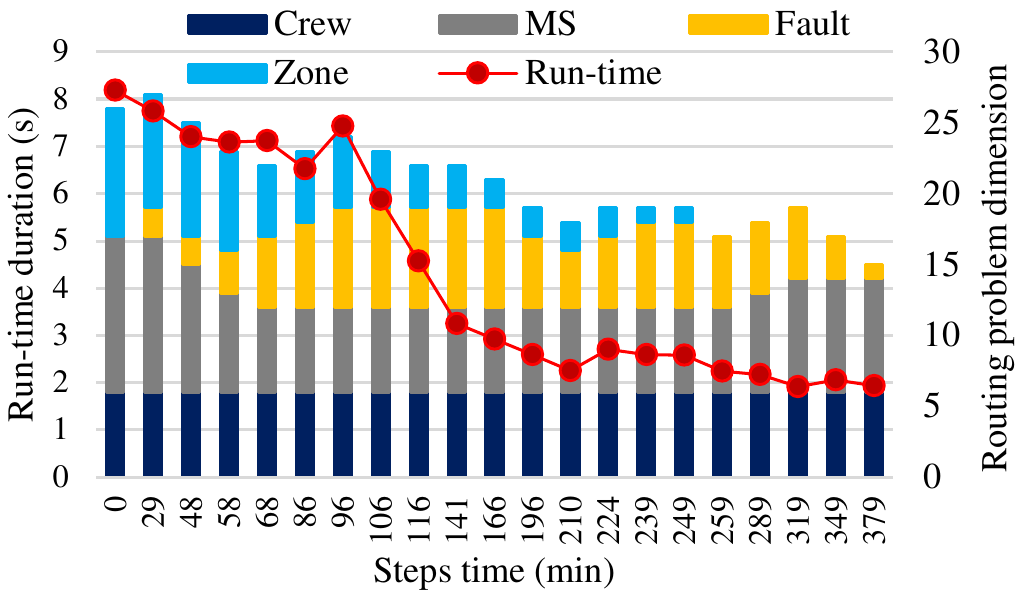}
\vspace{-1em}
\caption{Program run-time across different steps}
\label{fig: Program_run_time}
\vspace{-1em}
\end{figure}

Numerical simulations were conducted using Gurobi 8.1.1 on a system equipped with an AMD Ryzen7 4800H processor and 16 GB of memory. The model was found to be computationally efficient, with the optimization problem for all steps resolved in less than nine seconds, as depicted in Fig.~\ref{fig: Program_run_time}.

\begin{figure}[h]
\centering
\vspace{-1.25em}
\includegraphics[width=3.2in]{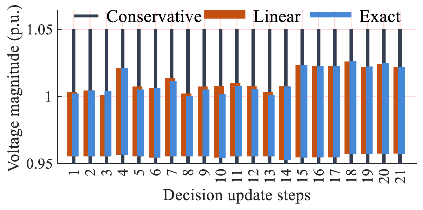}
\vspace{-1em}
\caption{Ranges of voltage magnitudes across time steps}
\label{fig: Voltage_levels}
\vspace{-1.5em}
\end{figure}

The complexity of the routing problem, as shown in Fig.~\ref{fig: Program_run_time}, is indicated by the number of crews, unpatrolled zones, faults, and the number of MS operations. According to the description of MSs operation in section~\ref{task assignment}, the number of potential MS operations is twice the number of closed MSs, as these could be opened and then reclosed, plus the number of opened MSs.
For the computation of precise minimum and maximum voltage levels, a multi-time step approach was employed, incorporating both linear and non-linear AC power flow constraints. However, the decision variables pertaining to the routing problem and the ultimate network configuration remained consistent with the time-step-free conservative scenario. In Fig.~\ref{fig: Voltage_levels}, the voltage magnitude ranges are depicted across three scenarios: the conservative time-step-free model with linear power flow, the multi-time step model with linear power flow, and the multi-time step model with non-linear exact power flow. As expected, the minimum and maximum values lie within the range of conservative bounds. Note that in the time-step-free model, the constraints merely enforce the upper and the lower bounds to be in the statutory range, allowing these variables to freely extend to the extreme ends. Therefore, our purpose here is not to assess the tightness of upper and lower bounds; an assessment of their tightness is deferred to section~\ref{subsec:Non-conservative Power Flow Approach}.

\subsection{Decision Update Frequency}\label{subsec:Decision_Update_Frequency}

\begin{figure}[h]
\vspace{-1em}
\centering
\includegraphics[width=3.2in]{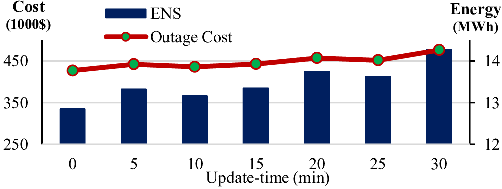}
\vspace{-1em}
\caption{Sensitivity of outage cost and ENS to decision update frequency}
\label{fig: Sensitivity_Analysis}
\vspace{-1.5em}
\end{figure}

In practical scenarios, as data regarding fault locations and repair times are progressively revealed through ongoing patrol operations, decision updates must be frequently performed to accommodate this newly acquired information. However, the immediacy of response to this new data is curtailed by factors such as data collection and processing time, as well as the runtime of various programs required for operations like load/generation estimation, travel time prediction, and fault management.
Fig.~\ref{fig: Sensitivity_Analysis} illustrates the sensitivity of the total network outage cost and energy not supplied (ENS) within the study horizon to variations in the minimum decision update time. (\textcolor{black}{In this case, we use the same proactive maximum update time as the base case, i.e., 30 minutes.}) A comparison of the results for update times of $5$ and $10$ minutes reveals that a more rapid response does not necessarily translate to cost reduction. This finding underscores the challenge of the exploration-exploitation dilemma in the context of dynamic decision-making in this environment.

\subsection{\textcolor{black}{Time-Step-Free Approach}}\label{subsec:Non-conservative Power Flow Approach}

\begin{figure}[h]
\centering
\includegraphics[width=3.in]{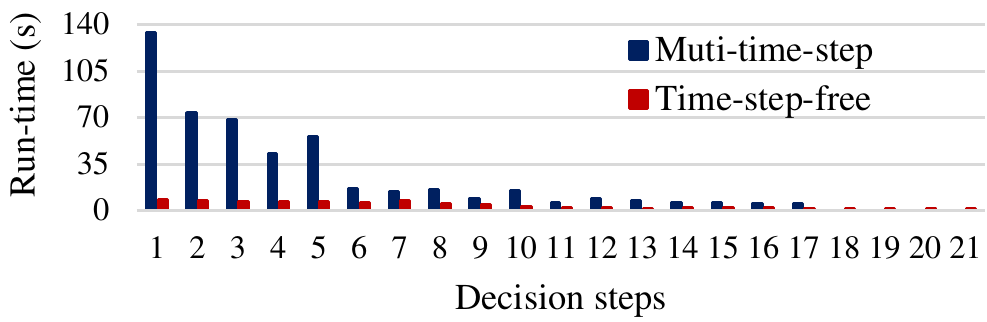}
\vspace{-1em}
\caption{\textcolor{black}{Computation times for time-step-free and multi-time-step approaches}}
\label{fig: Program_runtime_approaches}
\vspace{-1.5em}
\end{figure}

We next reassessed the base case scenario with the proposed methodology, replacing our conservative time-step-free power flow (PF) constraints with conventional multi-time-step linear PF constraints. This yields a marginal improvement (0.7\%) in the total network outage cost, from \$435.9k in the conservative power flow scenario to \$432.9k in the conventional scenario. The proximity of the outage costs showcases the tightness of our proposed bounds in this case. The run-times for each stage, for both the conventional and conservative approaches, are shown in Fig.~\ref{fig: Program_runtime_approaches}. While the two approaches suggest differing decisions and the problem parameters diverge after the initial stage, a clear uptick in overall computational complexity is observed when implementing conventional PF constraints.

\subsection{Simultaneous Restoration and Damage Assessment}\label{subsec:Simultaneous Restoration and Damage Assessment}

To benchmark the effectiveness of the proposed concurrent damage assessment and load restoration strategy, we considered two alternative benchmarks:

\begin{enumerate}[wide, labelwidth=!, labelindent=0pt]
\item \emph{First Patrol all, then Repair all (FPTR)}: Here, all crews are initially dispatched for feeder patrol and damage assessment. The objective at this stage is to minimize total patrol time \cite{lim2018}. Subsequently, fault repair is carried out to restore all loads.
\item \emph{Separate Patrol and Repair Crews (SPRC)}: In this scenario, crews $1$ and $5$ are assigned to patrol, while the others perform repairs. Fault repair is based on progressively updated information about the location and repair time of faults \cite{bian2022}.
\end{enumerate}

\begin{figure}[h]
\centering
\vspace{-0.7em}
\includegraphics[width=3.2in]{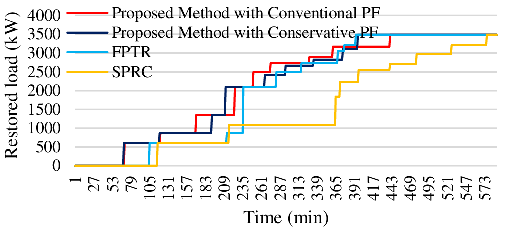}
\vspace{-1em}
\caption{Restored load over time in different restoration approaches}
\label{fig: Restored_load}
\vspace{-0.5em}
\end{figure}

\begin{figure}[h]
\centering
\includegraphics[width=3.2in]{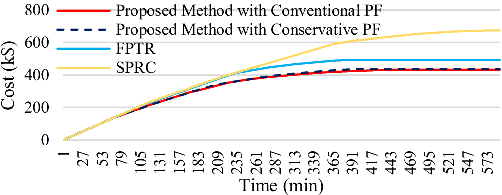}
\vspace{-1em}
\caption{Cumulative outage cost from the beginning of the process}
\label{fig: Cumulative_outage_cost}
\vspace{-1.5em}
\end{figure}

As Fig.~\ref{fig: Restored_load} demonstrates, our proposed method outperforms the others. The SPRC approach keeps repair crews idle until some faults are assessed. \textcolor{black}{While not leaving any crews idle, the FPTR approach shows lower performance than our proposed method since it prioritizes patrol actions over repair activities.} Fig.~\ref{fig: Cumulative_outage_cost} illustrates the cumulative outage cost from the beginning of the process.

\subsection{Scalability of the Solution Approach}\label{subsec:Scalability Solution Approach}

\begin{figure}[h]
\centering
\includegraphics[width=2.9in]{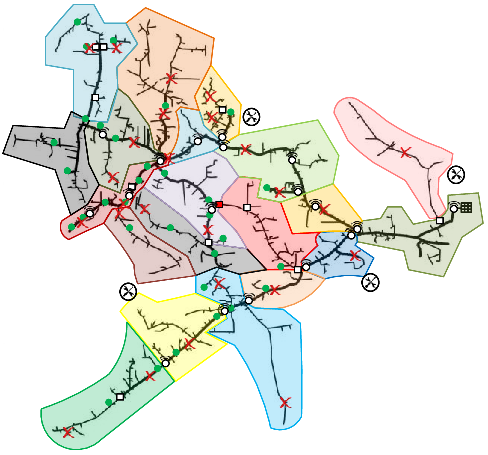}
\caption{Patrol zones in the IEEE 8500-node network}
\label{fig: Patrol_zones_8500}
\end{figure}

\begin{figure}[h]
\centering
\includegraphics[width=3.2in]{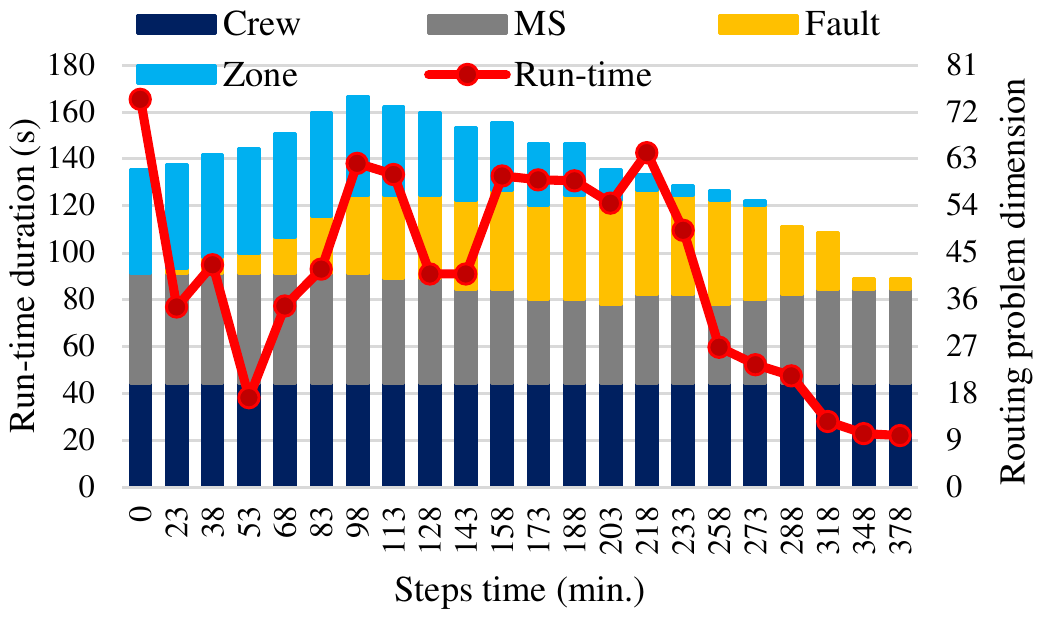}
\vspace{-1em}
\caption{Computation time and routing problem dimension across steps}
\label{fig: Program_runtime_routing_dimension}
\end{figure}

To assess the applicability of the proposed model for large-scale, real-world networks, we used the \texttt{IEEE 8500-node} system~\cite{8500-node}.
This network was partitioned into $20$ patrol zones, as depicted in Fig.~\ref{fig: Patrol_zones_8500}. We assumed that the network experienced a significant event, resulting in $25$ equipment damages. Within this network, $20$ crews, initially stationed at four locations, were tasked with damage assessment and service restoration. $32$ randomly placed DGs with random capacity from $100$ to $600$ kW are shown with green-filled circles.  
The computation time for all optimization steps was less than $170$ seconds, as shown in Fig.~\ref{fig: Program_runtime_routing_dimension}. This figure also reveals the routing problem's dimension, which includes the number of crews, unpatrolled zones, faults, and MSs operations. 
The results indicate that the proposed method offers an efficient and scalable solution for power system restoration, applicable even to large-scale networks.
\begin{figure}[ht]
\centering
\includegraphics[width=2.9in]{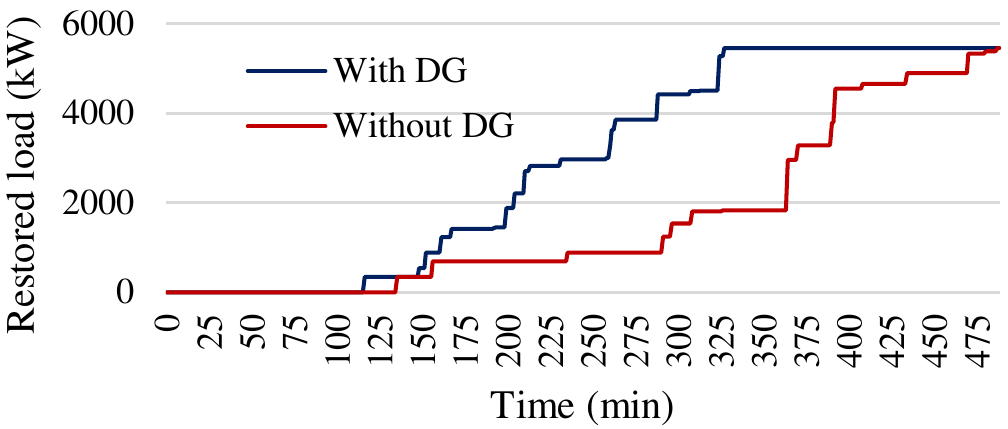}
\caption{DGs' impact on faster load restoration in 8500-bus test case}
\label{fig: DGs' impact}
\end{figure}
\textcolor{black}{ Additionally, to study how DGs facilitate faster load restoration (especially in cases with long feeders connected to a single substation), Fig.~\ref{fig: DGs' impact} illustrates the restored load over time for both the original case described above and that same case modified to remove the DGs. Observe that the restored power in the case without DGs significantly lags the case with DGs.}


\vspace{-0.5em}
\section{Conclusion}\label{sec: Conclusion}
This paper proposes a dynamic fault management plan designed for co-optimizing damage assessment and service restoration. The primary objective minimizes the total cost accrued from both outages and the restoration process. This objective is achieved by devising a routing plan for field crews, which includes feeder patrol, damage assessment, manual switching, and repair actions. To ensure the safe operation of the network in abnormal configurations, a conservative set of power flow equations is employed. This approach contributes to the efficiency and scalability of the proposed framework.

The results demonstrate the efficacy of simultaneous optimization and operation of feeder patrolling, damage assessment, repair, and restoration. By integrating these activities, significant benefits are observed in terms of outage reduction for the distribution network. This approach outperforms sequential phases or the deployment of separate crews for different actions. The analysis reveals that incorporating conservative power flow constraints can substantially alleviate the computational burden associated with the problem. Despite the reduced complexity, the total cost remains remarkably close to optimal levels. Consequently, the proposed fault management method holds promise for practical applicability in large-scale real-world distribution networks.

\textcolor{black}{We note that fault management is susceptible to data uncertainties in some parameters that we considered deterministically, such as the repair times provided by the damage assessors and the crews' travel times. Recognizing that mitigating these uncertainties could yield an even more effective decision-making process, our future work aims to formulate and solve problems which explicitly consider these uncertainties to enhance the overall adaptability of our methodology.}

\bibliographystyle{IEEEtran}

\bibliography{refs}

\ifarxiv
\textcolor{black}{\appendix[Passive and Active Loading]}
\textcolor{black}{\label{Passive and Active loading}}
\textcolor{black}{This appendix describes how the passive and active loading conditions result in upper and lower bounds on the voltage magnitudes as well as an upper bounds on the line flows, all of which are valid in the context of the linearized power flow model from~\cite{low2014pscc}.}

\newpage
\textcolor{black}{Consider a solution to the passive condition \eqref{eq: conservative PF} denoted as $\underline{p}^{*}=\left(\underline{U}^{*}, \underline{\varphi}^{*}, \underline{G}^{*}\right)$. If, for all buses $b$, $\underline{G}_{b}^{*} \geq G_{b}^{\min }$, then $\underline{p}^{*}$ can be deemed a feasible solution to~\eqref{eq: power flow cont} at the final time step, with $LB\{U\}=\min_ {b}\left\{\underline{U}_{b}^{*}\right\}$. Conversely, if some buses $b$ have $\underline{G}_{b}^{*} \leq G_{b}^{\min }$, we can remedy this by increasing generation at these buses until $G_{b}=G_{b}^{\min }$. This yields a new solution $p^{*}=\left(U^{*}, \varphi^{*}, G^{*}\right)$, and hence $U^{*} \geq \underline{U}^{*}$ due to to the increased power generation. Since this new $p^{*}$ satisfies~\eqref{eq: power flow cont} at the final time step, the aforementioned lower bound $LB\{U\}$ is still valid. Next, consider a solution to the active loading condition \eqref{eq:  active loading} denoted as $\overline{p}^{*}=\left(\overline{U}^{*}, \overline{\varphi}^{*}, \overline{G}^{*}\right)$. In this loading condition, we have $\overline{G} \geq G^{*}$ due to \eqref{eq: active loading:3} and \eqref{eq: active loading:4}, implying $\overline{U}^{*} \geq U^{*}$. Accordingly, if $\underline{U}^{*} \geq U^{\min }$ and $\overline{U}^{*} \leq U^{\max }$, as enforced in~\eqref{eq: Passive loading:5} and~\eqref{eq: active loading:6}, then $p^{*}=\left(U^{*}, \varphi^{*}, G^{*}\right)$ is a solution that satisfies \eqref{eq: power flow cont} at the final time step, thus confirming that the $\beta_{\ell}^{\text {line }}$ values correspond to a valid configuration at this time step.}

\textcolor{black}{Although the set of $\beta_{\ell}^{\text {line }}$ values only explicitly describe the final network configuration, the passive and active loading constraints are instrumental in validating previous steps. As we move backwards from step $t$ to $t-1$ by disconnecting the last-connected zone, we find that $\underline{U}^{*(t-1)} \geq \underline{U}^{*(t)}$ and $\overline{U}^{*(t-1)} \leq \overline{U}^{*(t)}$. Using similar reasoning, we can thus confirm the existence of a solution that satisfies \eqref{eq: power flow cont} for each previous time step.}

\textcolor{black}{Furthermore, we can infer that either $\left|\underline{\varphi}^{*}\right|$ or $\left|\overline{\varphi}^{*}\right|$ provides an upper bound for line flows at each step. To see this, note that, for each line at each step, if the state of the network is characterized by $p^{*}=\left(U^{*}, \varphi^{*}, G^{*}\right)$ and the line flow is directed downstream of the feeder, then $\left|\varphi_{\ell}^{*}\right| \leq\left|\underline{\varphi}_{\ell}^{*}\right|$. Since the network has a radial structure and there are no losses modeled in the power flow linearization from~\cite{low2014pscc}, the power flow on each line is determined by subtracting the total downstream power generation from the power consumption, i.e., $\underline{\varphi}_{\ell}^{*}=\sum_{b\in DS} (D_{b}-\underline{G}_{b}^{*})$, where $DS$ denotes the set of downstream buses to line $\ell$. Moreover, within the context of passive loading, all power generation levels are lower than the actual power generation, i.e., $\underline{G}^{*} \leq G^{*}$. Similarly, if the line flow is toward the upstream of the feeder, then $\left|\varphi_{\ell}^{*}\right| \leq\left|\overline{\varphi}_{\ell}^{*}\right|$ due to the conditions $\overline{\varphi}_{\ell}^{*}=\sum_{b\in DS} (\overline{G}_{b}^{*}-D_{b})$ and $\overline{G}^{*} \geq G^{*}$. Thus, feasibility of $\left|\underline{\varphi}^{*}\right|$ or $\left|\overline{\varphi}^{*}\right|$, as enforced in~\eqref{eq: Passive loading:5} and~\eqref{eq: active loading:6}, ensures that the line flow limits are satisfied for the final time step. Using a similar argument as in the case of the voltage limits, the current limits are also satisfied for all previous time steps. Thus, enforcing feasibility for the passive and active loading conditions in the final time step ensures that both the voltage and current limits will be satisfied for all time steps, in the context of the linearized power flow model from~\cite{low2014pscc} as in~\eqref{eq: power flow cont}.}

\else
\fi

\end{document}